\def\crb{CrB$_2$}
\def\mgb{MgB$_2$}

\def\a2f{$\alpha^2F$}
\def\transa2f{$\alpha^2_{tr}F$}

\documentclass[aps,prl,twocolumn, superscriptaddress,longbibliography]{revtex4-2}

\usepackage{graphicx}
\usepackage{pstricks}
\usepackage{amsmath}
\usepackage{gensymb}
\usepackage{amsfonts}
\usepackage{dcolumn}
\usepackage{bm}
\usepackage{makecell}
\usepackage{xcolor}
\usepackage{float}
\usepackage{xcolor}
\usepackage{hyperref}

\makeatletter
\AtBeginDocument{\let\LS@rot\@undefined}
\makeatother

\begin{document}
\title{Hybrid $s$-wave superconductivity in \crb }
\author{Sananda Biswas}
\email{biswas@itp.uni-frankfurt.de}
\affiliation{Institut f\"ur Theoretische Physik, Goethe-Universit\"at Frankfurt, 60438 Frankfurt am Main, Germany}
\author{Andreas Kreisel}
\email{kreisel@itp.uni-leipzig.de}
\affiliation{Institut  f\"ur Theoretische Physik, Universit\"at Leipzig, Br\"uderstr. 16, 04103 Leipzig, Germany}
\affiliation{Niels Bohr Institute, University of Copenhagen, 2100 Copenhagen, Denmark}
\author{Adrian Valadkhani}
\affiliation{Institut f\"ur Theoretische Physik, Goethe-Universit\"at Frankfurt, 60438 Frankfurt am Main, Germany}
\author{Matteo D\"urrnagel}
\affiliation{Julius-Maximilians-Universit\"at W\"urzburg, W\"urzburg, Germany}
\affiliation{Institute for Theoretical Physics, ETH Z\"{u}rich, 8093 Z\'{u}rich, Switzerland}
\author{Tilman Schwemmer}
\author{Ronny Thomale}
\affiliation{Julius-Maximilians-Universit\"at W\"urzburg, W\"urzburg, Germany}
\author{Roser Valent\'i}
\affiliation{Institut f\"ur Theoretische Physik, Goethe-Universit\"at Frankfurt, 60438 Frankfurt am Main, Germany}
\author{Igor I. Mazin}
\affiliation{Department of Physics and Astronomy, George Mason University, Fairfax, VA 22030}
\affiliation{Quantum Science and Engineering Center, George Mason University, Fairfax, VA 22030}

\begin{abstract} 
In a metal with multiple Fermi pockets, the formation of $s$-wave superconductivity can be conventional due to electron-phonon coupling or unconventional due to spin fluctuations. We analyze the hexagonal diboride \crb, which is an itinerant antiferromagnet at ambient conditions and turns superconducting upon increasing pressure. While the high pressure behavior of $T_c$ suggests conventional $s$-wave pairing, we find that spin fluctuations promoting unconventional $s$-wave pairing become important in the vicinity of the antiferromagnetic dome. As the symmetry class of the $s$-wave state is independent of its underlying mechanism, we argue that \crb\ is a realization of a {\it hybrid} s-wave superconductor where unconventional and conventional s-wave mechanisms team up to form a joint superconducting dome.
\end{abstract}
\date{\today}
\maketitle

{\it Introduction.-} 
Even though the phenomenological description of a superconducting state  finds its common ground in the notion of a phase-coherent superposition of Cooper pairs and the mean field  description derived thereof~\cite{BCS}, the possible microscopic formation mechanism is highly diverse. In a so-called conventional superconductor (CS), electron-phonon coupling generates an effective attractive electron-electron interaction~\cite{eliashberg,PhysRev.117.648,PhysRev.125.1263}. Not only do phonons promote zero-angular-momentum Cooper pairs, $i.e.$, an $s$-wave type pairing function, but the electron-phonon interaction  also tends to be relatively momentum-independent, a few exceptions not withstanding, and gives rise to a reasonably uniform gap, $\Delta_{\text{CS}}({\bf{k}})\sim \text{const.}$, throughout the Brillouin zone. 

\begin{figure}[!hb]
\includegraphics[width=0.8\columnwidth]{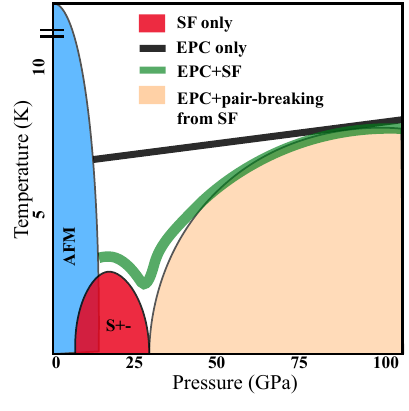}
\caption{Proposed schematic temperature-pressure phase diagram for a hybrid $s$-wave superconductor. Spin fluctuation limit:  Antiferromagnetic region with maximum $T_{\text{AFM}}$ at ambient pressure (blue region), dropping rapidly with pressure and followed by (possibly overlapping with) a smaller unconventional superconducting 
dome near the quantum critical point (red region).  Electron-phonon limit: $T_c$ is only weakly dependent on pressure and dominates the high-pressure part of the phase diagram (light orange region). The hybrid $s$-wave appears in the crossover region between the two limits where the pair breaking impact of spin fluctuations is vital for explaining the drop of $T_c$ towards ambient pressure.}
\label{fig1}
\end{figure}

For unconventional superconductivity (UCS), which has most prominently surfaced in the context of the high-$T_c$ cuprates, the microscopic footing of pairing appears both more diverse and less understood. From the viewpoint of spin fluctuations  ~\cite{PhysRevLett.15.524}, electron pairing can originate from repulsive electron-electron interactions~\cite{gros1987,white1989}. However, this implies that the gap function is sign-changing in the Brillouin zone: $\langle \Delta_{\text{UCS}}({\bf{k}})\rangle_{\text{BZ}}^2\ll \langle \Delta^2_{\text{UCS}}({\bf{k}})\rangle_{\text{BZ}} $. For a single pocket Fermi surface, this naturally suggests the presence of nodes and thus a principal unconventional superconducting gap that is qualitatively different from the conventional one. For multiple Fermi pockets, however, nodes are avoidable for unconventional pairing by allowing sign changes of $\Delta$ between the pockets. This is at the heart of the nature of superconducting pairing in iron pnictide superconductors~\cite{MazinPRL2008,hirschfeld2011}, where the compensated metal parent state forms superconducting pairing of opposite sign for hole and electron pockets, respectively. For each individual pocket, the gap may appear rather uniform, even though typically not as uniform as for a conventional superconducting state. Note that from the viewpoint of symmetry, such an unconventional superconducting gap cannot be distinguished from a conventional $s$-wave: in both cases, the zero momentum Cooper pair is described by the irreducible point group representation with trivial characters. Yet, even though the {\it angular} dependence of the superconducting gap  as a function of the wave vector is the same, the {\it radial} part shifts away from approximately constant in the conventional superconductor, referred to as $s_{++}$, to exhibit strong $k$-dependence in the unconventional case, which is referred to as $s_{+-}$.

Since an $s$-wave superconductor can exists both as a conventional $s_{++}$ and and unconventional $s_{+-}$, in principle, a 
{\it hybrid $s$-wave superconductor} can be imagined, where unconventional and conventional pairing mechanisms team up to yield one continuous $s$-wave superconducting region spanning both CS and UCS domains. This intriguing possibility, however, requires a hypothetical material where spin fluctuations and phonons have a possibility to cooperate at least to some extent.  More specifically, the latter have to peak at small wave vector, while the former (as long as we are not considering triplet pairing) necessarily have to have a maximum at a particular wave vector matching the Fermi surface geometry.

While iron pnictide superconductors seem to be a promising host for a hybrid superconducting state, the conventional and unconventional pairing are too imbalanced because of good Fermi surface nesting enhancing the spin fluctuation (SF) mechanism, as compared to weak electron-phonon coupling (EPC)~\cite{PhysRevLett.101.026403}. In addition, EPC does not satisfy the requirement to peak at small $q$ ($i.e.$, in the intraband channel) either. In fact, the competition between SF and EPC has been under scrutiny for quite a few superconducting materials~\cite{ybco_nunner,cubiso_boeri,dolgov_mgcni3, moiresuperlattice,Amy}, but no suitable candidate has been identified so far, exhibiting not only competition, but also collaboration between the two mechanisms in some range of parameters.

In this Letter, we propose \crb\ as a potential  hybrid $s$-wave superconductor, continuously tunable by pressure between the  $s_{+-}$ and $s_{++}$ limits. Indeed, superconductivity has recently been discovered in \crb\ under pressure, with a maximum $T_c$ of 7 K~\cite{pei2021pressureinduced}. Isostructural to the conventional superconductor \mgb~\cite{nagamatsu2001, Mazin_mgb2}, \crb\ exhibits itinerant antiferromagnetism at ambient pressure. $T_c$ is found to weakly increase with pressure at odds with the typical dome formation in unconventional superconductors. This is a strong indication that at least beyond a certain pressure regime, superconducting pairing in \crb\ is dominated by the electron-phonon coupling. Due to the proximity to the antiferromagnetic order, however, the superconductivity in \crb\ is likely to be of unconventional nature at lower pressure, since spin fluctuations are crucial near a magnetic quantum critical point. From our synoptic analysis of superconductivity, accounting for both electron-phonon coupling and spin fluctuations, we find that \crb\ is a likely candidate for a hybrid $s$-wave superconductor, as summarized in Fig.~\ref{fig1}. On the one hand, spin fluctuations promoting $s_{+-}$-type are highly relevant due to significant nesting of the \crb\ Fermi surface (note that here the three-dimensional nesting is present due to the flat parts in the Fermi surface in the $k_x-k_y$ plane with large density of states). On the other hand, our calculations yield rather strong electron-phonon coupling, with the main contribution coming from around the Brillouin-zone center, $\Gamma$, enabling a significant cooperative effect between the two mechanisms. The interplay of both pairing tendencies culminates in the schematic temperature-pressure phase diagram depicted in Fig.~\ref{fig1} (green line): The superconducting domain in the high pressure limit is dominated by the electron-phonon mechanism, which, if let alone, would suggest a fairly constant $T_c$ throughout the phase diagram (black line) with no indication for competitive orders. By contrast, only  spin fluctuations related to the antiferromagnetic order at the ambient pressure would generate unconventional superconductivity within a small pressure range and a significantly lower $T_c$ than $T_{\text{AFM}}$ (red dome). Taken together, however, electron-phonon couplings and spin fluctuations suggest a crossover between the two limiting scenarios, where, starting from the center of the superconducting region, spin fluctuations act upon the $s$-wave superconductor as pair breakers close to the antiferromagnetic phase at lower pressure and the electron-phonon coupling determines the scale of saturating superconducting pairing for higher pressure.

\begin{figure}[!b]
\centering
\includegraphics[width=1\columnwidth]{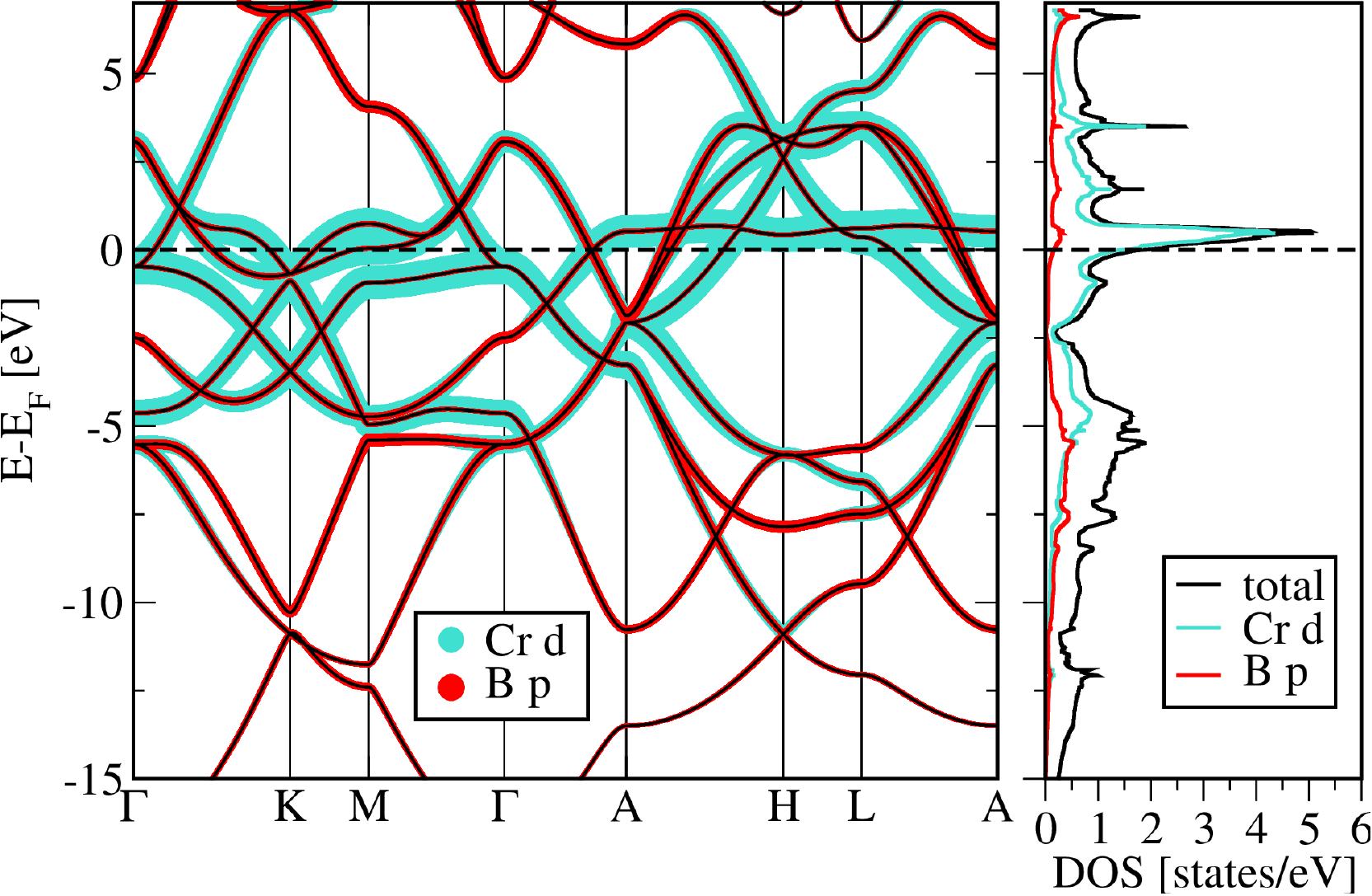}
\caption{Electronic band structure and density of states (DOS) at 100 GPa. Fat bands and DOS of Cr $3d$ states (cyan) and B $2p$ states (red) show that both states contribute to the formation of the Fermi surface.}
\label{fig:bands}
\end{figure}

{\it Methods.-} 
First, we investigate the superconducting pairing from a pure electron-electron interaction perspective adapting the spin-fluctuation pairing mechanism expected to be present for the correlated Cr $3d$ orbitals. To this end, we set up the pairing interaction in the random phase approximation (RPA)~\cite{Graser2009,altmeyer2016} and solve the linearized gap equation for a discretization of the three dimensional Fermi surface~\cite{Kreisel2013,Durrnagel2022, supplemental}, yielding the eigenvalues, $\lambda_i$ and the gap eigenfunctions, $g_i(\bf{k})$, proportional to the superconducting order parameter. Next, we examine the electron-phonon mechanism by calculating the pressure dependence of the phonon dispersion, electron-phonon line-width, $\gamma$ and electron-phonon coupling constants, $\lambda$, using density functional perturbation theory as implemented in the Quantum ESPRESSO code \cite{QE}.

\begin{figure}[!t]
\centering
\includegraphics[width=\columnwidth]{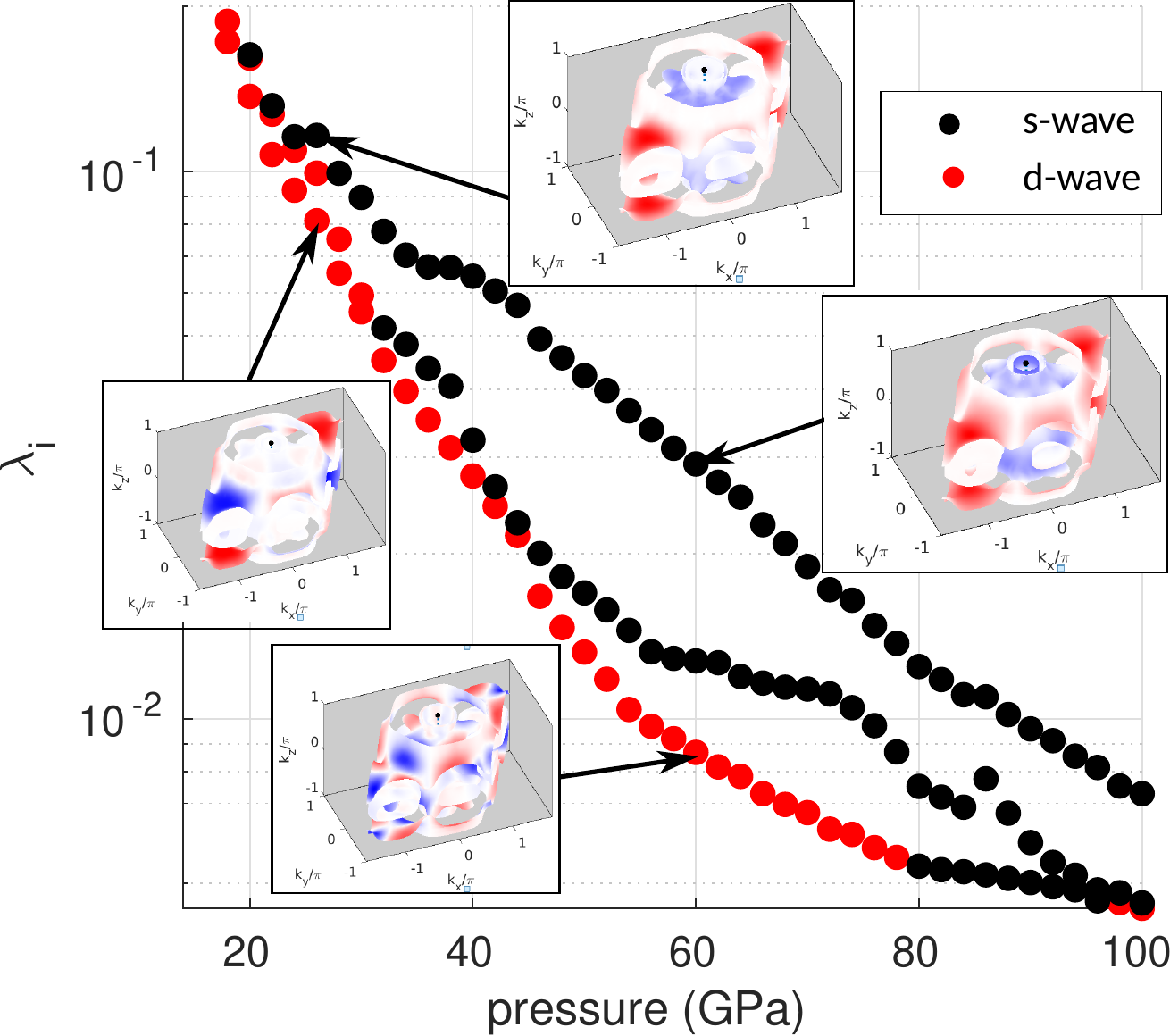}
\caption{Spin fluctuation pairing. Eigenvalues $\lambda_i$ of the leading and sub-leading instabilities as a function of pressure as calculated with $U=0.124\;\text{eV}$ where the critical pressure is $p_c=16\;\text{GPa}$. Insets: Gap function $g_i(\mathbf{k})$ on the Fermi surface of the leading $s$-wave (black circles) state and $d$-wave (red circles) state for two representative pressures as indicated by the arrows. Red and blue regions represent two opposite signs of the gap function.}
\label{fig_mgb2PH}
\end{figure}

{\it Results.-} 
\crb\ crystallizes in a $P6/mmm$ structure (space group 191)  with $c/a > 1$ at ambient pressure. As a function of pressure the $c$ parameter decreases more rapidly than $a$ and, around 30 GPa the ratio $c/a$ becomes less than 1 with no observed changes in the crystal structure \cite{pei2021pressureinduced}. While the Mg states in \mgb\ lie far away from the Fermi surface, the band structure of \crb\  suggests significant contributions from Cr $3d$ orbitals near the Fermi surface at all pressures (See Fig.~\ref{fig:bands} for the 100 GPa case), along with  contributions coming from B atoms~\cite{supplemental}. This makes the properties of \crb\ markedly different from the isostructural \mgb \cite{kortus2001}, as we will discuss below.  Accounting for the spin-density wave instability at low pressures~\cite{bauer, pei2021pressureinduced}, we tune the bare Coulomb interaction responsible for the spin-fluctuations such that the RPA instability occurs roughly at $p_c=$ 16 GPa  and then calculate pairing eigenvalues $\lambda_i$ as a function of pressure. Fig. \ref{fig_mgb2PH} reveals the two main results from this investigation: First, the eigenvalues $\lambda_i$ and therefore also the pairing strength rapidly decrease as a function of pressure, making this pairing interaction practically irrelevant at large pressures. Second, the leading instability is of sign changing $s_{+-}$  type, with one sign at a band forming a flat part at finite $k_z$ close to the Brillouin zone boundary and opposite sign at the Fermi surface appearing around the $\Gamma$ point (see  Supplemental Material~\cite{supplemental} for top views).  Higher order $s$-wave solutions or $d$-wave solutions are sub-leading and their eigenvalues exhibit peaks from van Hove singularities moving through the Fermi level.

\begin{figure}[!t]
\centering
\includegraphics[width=1\columnwidth]{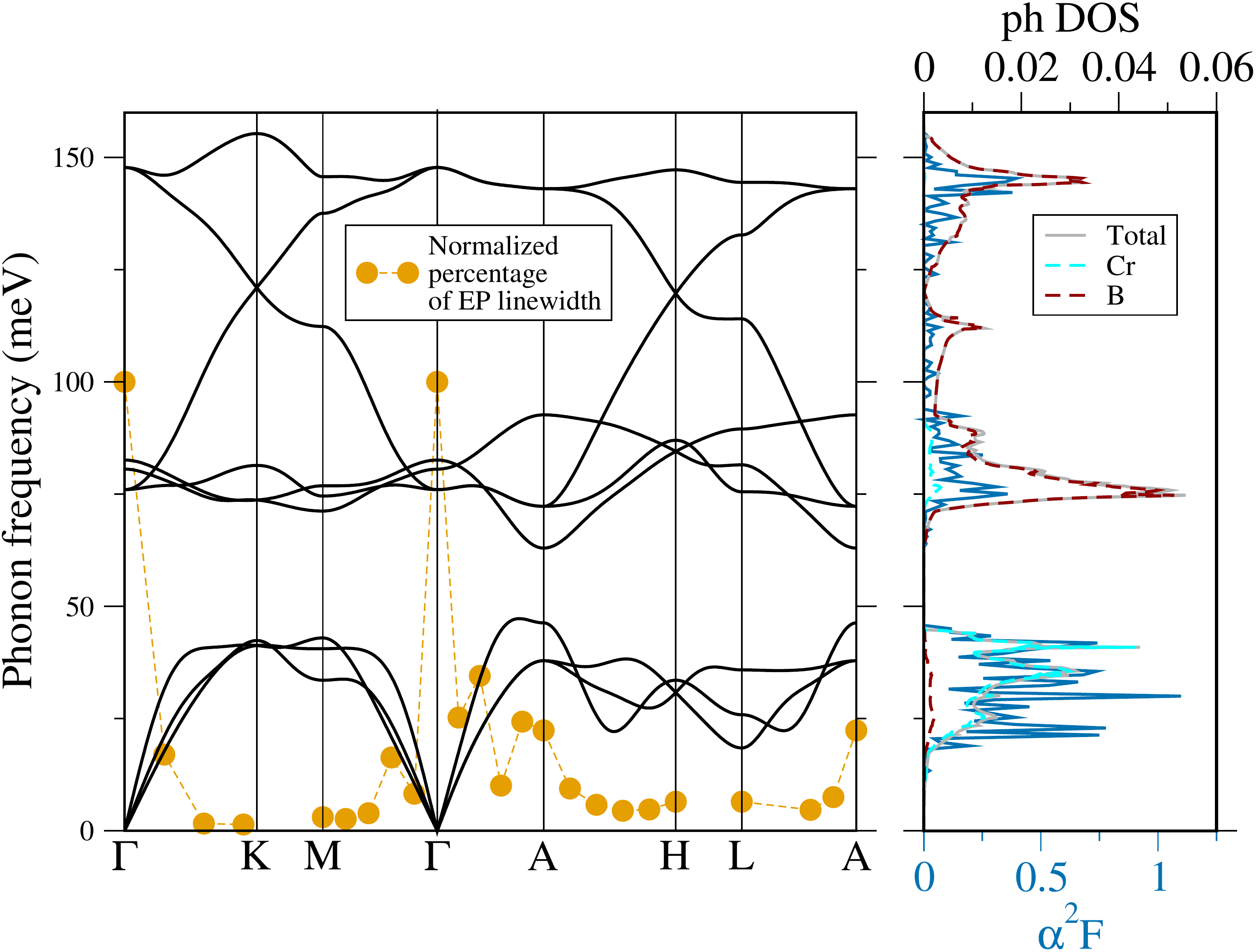}
\caption{Phonon dispersion, electron-phonon  (EP) linewidth, spectral function [$\alpha^2F(\omega)$], and phonon density of states (ph DOS)  of the relaxed structure at 100 GPa. Normalized percentage of EP line-width (orange) is equal to $\gamma$/$\gamma_{\mathrm{max}}\times 100\%$ and is strongly peaked at zone center, $\Gamma$.}
\label{fig:3}
\end{figure}

In order to examine the electron-phonon pairing and the expected critical temperature from a material-specific perspective, we first determine the strength of the electron-phonon coupling constant, $\lambda$. Experimentally, the system is reported to have the largest $T_\mathrm c$ around 100 GPa; therefore, we first focus on this pressure region. The relaxed structure at 100 GPa has a ratio $c/a = 0.93$ (whereas $(c/a)_{\text{exp}}=0.96$) with all phonon modes being stable. We therefore have chosen this structure for establishing convergence criteria for $\lambda$ with respect to $\mathbf{k}$-mesh, $\mathbf{q}$-mesh and Gaussian broadening (see Ref.~\cite{supplemental}). The converged value of $\lambda = 0.78$ has significant contributions to the electron-phonon spectral function, $\alpha^2F(\omega)$, from the low-frequency vibrations involving Cr atoms as seen in Fig.~\ref{fig:3}. Note that the phonon dispersions calculated with the experimental structure~\cite{pei2021pressureinduced} exhibit an imaginary acoustic mode along the $\mathbf{k}$-path perpendicular to the honeycomb boron plane. Though this indicates an instability toward a charge density wave state, zero-point vibrational effects can stabilize the conventional structure~\cite{supplemental}.  Nonetheless, an estimate of the lower    bound for $\lambda =0.60$ could be obtained by excluding the imaginary acoustic mode at 100 GPa.

The electron-phonon line-width, $\gamma$ (plotted as $\gamma$/$\gamma_{\mathrm{max}}\times 100\%$) exhibits maximum contribution around the Brillouin-zone center, $i.e.$, ${\bf q}=0$, which suggests significantly stronger intraband electron-phonon coupling compared to interband  contribution, giving rise to conventional $s_{++}$ pairing. With the average phonon frequency softening at lower pressures, $T_c$ is also reduced as pressure goes down. Our calculations at two values of pressure support this interpretation and justifies our sketch of the  electron-phonon driven contribution to $T_c$ in Fig.~\ref{fig1}.

{\it Discussions.-}
We have so far investigated the two pairing mechanisms individually, finding that the attractive electron-phonon pairing interaction is dominated by contributions close to ${\bf q}=0$, thus exhibiting a large and positive intraband attractive interaction $V^{\text{EPC}}_{\text{intra}}>0$, while the interband interaction is much smaller, $V^{\text{EPC}}_{\text{inter}}\ll V^{\text{EPC}}_{\text{intra}}$, leading to critical temperatures of $T_\mathrm{c}=$ 7 K. The pairing interaction from SF is generically repulsive (negative), but more repulsive for large momentum transfer so that the interband pairing $V^{\text{SF}}_{\text{inter}}<0$ is dominating over the intraband pairing  $|V^{\text{SF}}_{\text{inter}}|\gg |V^{\text{SF}}_{\text{intra}}|$.

Close to the spin-density-wave instability, the spin fluctuations are enhanced, leading to a sizable pairing interaction which then quickly decreases with pressure beyond the critical pressure $p_c$ (pressure at which spin fluctuations diverge). Assuming the usual RPA mechanism, one can trace this back to the increase of the electronic bandwidth as a function of pressure, $W(p)\approx W_{\text{AFM}}[1+x(p-p_c)]$ where $W_{\text{AFM}}$ is the bandwidth at the critical pressure $p_c$ and the dependence is assumed to be expanded to first order close to $p_c$. Taking this into account, we obtain $V^{\text{SF}}_{\text{intra/inter}}\propto [\alpha_{\text{intra/inter}}+x(p-p_c)]^{-1}$, where $\alpha_{\text{intra/inter}}$ describes the closeness of the intra (inter) band scattering vectors to a nesting vector giving rise to a peak in the susceptibility~\cite{supplemental}. 

Adding up the electron-phonon and the spin fluctuations pairing interactions, one then arrives at the qualitative behavior of the critical temperature depicted in Fig. \ref{fig1}, where the spin fluctuations dominate close to $p_c$ and a sign-changing order parameter emerges as the dominant instability. At larger pressures, the critical temperature is expected to decrease until the $s_{+-}$ and $s_{++}$ instabilities have comparable eigenvalues and the critical temperature increases again, since now the pair-breaking contribution from the spin-fluctuations $V^{\text{SF}}_{\text{intra}}$ decreases as well and the sum $V^{\text{EPC}}_{\text{intra}}+V^{\text{SF}}_{\text{intra}}$ is dominated by the electron-phonon contribution. Note that despite the small eigenvalues for the SF pairing instability, the electronic contribution to pairing remains finite due to the momentum-independent Hubbard-Hund interaction, and pair-breaking will remain non-negligible even at large pressures, where the expected $T_c$ for the `SF alone' pairing is exponentially small.

Even though isostructural, \crb\ therefore differs from the prototypical electron-phonon superconductor \mgb, due to the following aspects: First, due to the presence of  Cr 3$d$ states, additional effects from these correlated states contribute to the Fermiology of the system whereas Mg states in \mgb\ do not take part in the formation of Cooper pairs.  Second, the isotropic electron-phonon coupling constants are comparable ($\lambda|_{\text{\crb}(\mathrm{100 \ GPa})}=0.78$ and $\lambda|_{\text{\mgb}(\mathrm{P=0)}} =0.71$  ); nonetheless, unlike \mgb,\ the low-frequency Cr-phonon modes have significant contributions to the EP-spectral function.

{\it Conclusions.-} 
In summary, we have presented a hybrid perspective of electron-phonon  and spin-fluctuation pairing in order to explain the overall phase diagram of \crb{} where $T_c$ is found to increase at pressures far away from the antiferromagnetic instability, a behavior not expected for superconductors driven by spin fluctuations alone. Instead, the cooperative and anti-cooperative effects of the two pairing mechanisms allowed by the presence of a leading instability of the same symmetry explains the larger critical temperature at high pressure (far away from the 
antiferromagnetic order). Furthermore, the pairing state, while lowering pressure, is predicted to have a crossover from $s_{++}$ to $s_{+-}$ on approaching the quantum critical point with a non-monotonous behavior of $T_c$. We have discussed in detail the differences between the well studied  MgB$_2$ and the CrB$_2$ system; the latter has correlated $d$ states close to the Fermi level and a dominating electron-phonon interaction at small momentum transfer; both of them are required ingredients for the appearance of the hybrid $s$-wave superconductivity which is expected to exhibit an anisotropic order parameter due to unconventional pairing. Experimental signatures of the proposed scenario would be the non-monotonous behavior of $T_c$ close to the quantum critical point and the crossover to non-sign changing order parameter with increasing pressure. The latter could be tested experimentally by observing the effect of disorder. At lower pressures the potential scatterers should lead to a strong suppression of $T_c$, while the $s_{++}$ state at higher pressures should be almost unaffected according to Anderson's theorem. The ability to not only tune $T_c$~\cite{Basov2011}, but also the nature of the superconducting order parameter may open new perspectives in the study of unconventional superconductivity.

\begin{acknowledgments}
We thank Young-Joon Song, Paul Wunderlich and Shinibali Bhattacharyya for discussions. S.B., A.V. and R.V. thank the Deutsche Forschungsgemeinschaft (DFG, German Research Foundation) through TRR 288-422213477 (Projects A05, B05). A.K. acknowledges support by the Danish National Committee for Research Infrastructure (NUFI) through the ESS-Lighthouse Q-MAT. R.T and R.V. acknowledge support from the DFG through QUAST FOR 5249-449872909 (Projects P3, P4). I.I.M. acknowledges support from the U.S. Department of Energy through Grant no. DE-SC0021089 and from the Wilhelm and Else Heraeus Foundation. M.D., T.S. and R.T. acknowledge funding by the DFG through Project-ID 258499086 - SFB 1170 and through the W\"{u}rzburg-Dresden Cluster of Excellence on Complexity and Topology in Quantum Matter – ct.qmat Project-ID 390858490 - EXC 2147.
\end{acknowledgments}

% command such that additional references appear in the bibliography

\bibliography{crb2}

\begin{thebibliography}{31}%
\makeatletter
\providecommand \@ifxundefined [1]{%
 \@ifx{#1\undefined}
}%
\providecommand \@ifnum [1]{%
 \ifnum #1\expandafter \@firstoftwo
 \else \expandafter \@secondoftwo
 \fi
}%
\providecommand \@ifx [1]{%
 \ifx #1\expandafter \@firstoftwo
 \else \expandafter \@secondoftwo
 \fi
}%
\providecommand \natexlab [1]{#1}%
\providecommand \enquote  [1]{``#1''}%
\providecommand \bibnamefont  [1]{#1}%
\providecommand \bibfnamefont [1]{#1}%
\providecommand \citenamefont [1]{#1}%
\providecommand \href@noop [0]{\@secondoftwo}%
\providecommand \href [0]{\begingroup \@sanitize@url \@href}%
\providecommand \@href[1]{\@@startlink{#1}\@@href}%
\providecommand \@@href[1]{\endgroup#1\@@endlink}%
\providecommand \@sanitize@url [0]{\catcode `\\12\catcode `\$12\catcode
  `\&12\catcode `\#12\catcode `\^12\catcode `\_12\catcode `\%12\relax}%
\providecommand \@@startlink[1]{}%
\providecommand \@@endlink[0]{}%
\providecommand \url  [0]{\begingroup\@sanitize@url \@url }%
\providecommand \@url [1]{\endgroup\@href {#1}{\urlprefix }}%
\providecommand \urlprefix  [0]{URL }%
\providecommand \Eprint [0]{\href }%
\providecommand \doibase [0]{https://doi.org/}%
\providecommand \selectlanguage [0]{\@gobble}%
\providecommand \bibinfo  [0]{\@secondoftwo}%
\providecommand \bibfield  [0]{\@secondoftwo}%
\providecommand \translation [1]{[#1]}%
\providecommand \BibitemOpen [0]{}%
\providecommand \bibitemStop [0]{}%
\providecommand \bibitemNoStop [0]{.\EOS\space}%
\providecommand \EOS [0]{\spacefactor3000\relax}%
\providecommand \BibitemShut  [1]{\csname bibitem#1\endcsname}%
\let\auto@bib@innerbib\@empty
%</preamble>
\bibitem [{\citenamefont {Bardeen}\ \emph {et~al.}(1957)\citenamefont
  {Bardeen}, \citenamefont {Cooper},\ and\ \citenamefont {Schrieffer}}]{BCS}%
  \BibitemOpen
  \bibfield  {author} {\bibinfo {author} {J.~Bardeen}, \bibinfo {author} {L.~N.
  Cooper}\ and\ \bibinfo {author} {J.~R. Schrieffer},\ }\bibfield  {title}
  {\bibinfo {title} {Theory of superconductivity},\ }\href
  {https://doi.org/10.1103/PhysRev.108.1175} {\bibfield  {journal} {\bibinfo
  {journal} {Phys. Rev.}\ }\textbf {\bibinfo {volume} {108}},\ \bibinfo {pages}
  {1175} (\bibinfo {year} {1957})}\BibitemShut {NoStop}%
\bibitem [{\citenamefont {Eliashberg}(1960)}]{eliashberg}%
  \BibitemOpen
  \bibfield  {author} {\bibinfo {author} {G.~M. Eliashberg},\ }\href@noop {}
  {\bibfield  {journal} {\bibinfo  {journal} {Zh. Eksp. Teor. Fiz.}\ }\textbf
  {\bibinfo {volume} {38}},\ \bibinfo {pages} {966} (\bibinfo {year}
  {1960})}\BibitemShut {NoStop}%
\bibitem [{\citenamefont {Nambu}(1960)}]{PhysRev.117.648}%
  \BibitemOpen
  \bibfield  {author} {\bibinfo {author} {Y.~Nambu},\ }\bibfield  {title}
  {\bibinfo {title} {Quasi-particles and gauge invariance in the theory of
  superconductivity},\ }\href {https://doi.org/10.1103/PhysRev.117.648}
  {\bibfield  {journal} {\bibinfo  {journal} {Phys. Rev.}\ }\textbf {\bibinfo
  {volume} {117}},\ \bibinfo {pages} {648} (\bibinfo {year}
  {1960})}\BibitemShut {NoStop}%
\bibitem [{\citenamefont {Morel}\ and\ \citenamefont
  {Anderson}(1962)}]{PhysRev.125.1263}%
  \BibitemOpen
  \bibfield  {author} {\bibinfo {author} {P.~Morel}\ and\ \bibinfo {author}
  {P.~W. Anderson},\ }\bibfield  {title} {\bibinfo {title} {Calculation of the
  superconducting state parameters with retarded electron-phonon interaction},\
  }\href {https://doi.org/10.1103/PhysRev.125.1263} {\bibfield  {journal}
  {\bibinfo  {journal} {Phys. Rev.}\ }\textbf {\bibinfo {volume} {125}},\
  \bibinfo {pages} {1263} (\bibinfo {year} {1962})}\BibitemShut {NoStop}%
\bibitem [{\citenamefont {Kohn}\ and\ \citenamefont
  {Luttinger}(1965)}]{PhysRevLett.15.524}%
  \BibitemOpen
  \bibfield  {author} {\bibinfo {author} {W.~Kohn}\ and\ \bibinfo {author}
  {J.~M. Luttinger},\ }\bibfield  {title} {\bibinfo {title} {New mechanism for
  superconductivity},\ }\href {https://doi.org/10.1103/PhysRevLett.15.524}
  {\bibfield  {journal} {\bibinfo  {journal} {Phys. Rev. Lett.}\ }\textbf
  {\bibinfo {volume} {15}},\ \bibinfo {pages} {524} (\bibinfo {year}
  {1965})}\BibitemShut {NoStop}%
\bibitem [{\citenamefont {Gros}\ \emph {et~al.}(1987)\citenamefont {Gros},
  \citenamefont {Joynt},\ and\ \citenamefont {Rice}}]{gros1987}%
  \BibitemOpen
  \bibfield  {author} {\bibinfo {author} {C.~Gros}, \bibinfo {author}
  {R.~Joynt}\ and\ \bibinfo {author} {T.~Rice},\ }\bibfield  {title} {\bibinfo
  {title} {Superconducting instability in the large-u limit of the
  two-dimensional hubbard model},\ }\href@noop {} {\bibfield  {journal}
  {\bibinfo  {journal} {Zeitschrift f{\"u}r Physik B Condensed Matter}\
  }\textbf {\bibinfo {volume} {68}},\ \bibinfo {pages} {425} (\bibinfo {year}
  {1987})}\BibitemShut {NoStop}%
\bibitem [{\citenamefont {White}\ \emph {et~al.}(1989)\citenamefont {White},
  \citenamefont {Scalapino}, \citenamefont {Sugar}, \citenamefont {Bickers},\
  and\ \citenamefont {Scalettar}}]{white1989}%
  \BibitemOpen
  \bibfield  {author} {\bibinfo {author} {S.~White}, \bibinfo {author}
  {D.~Scalapino}, \bibinfo {author} {R.~Sugar}, \bibinfo {author} {N.~Bickers}\
  and\ \bibinfo {author} {R.~Scalettar},\ }\bibfield  {title} {\bibinfo {title}
  {Attractive and repulsive pairing interaction vertices for the
  two-dimensional hubbard model},\ }\href@noop {} {\bibfield  {journal}
  {\bibinfo  {journal} {Physical Review B}\ }\textbf {\bibinfo {volume} {39}},\
  \bibinfo {pages} {839} (\bibinfo {year} {1989})}\BibitemShut {NoStop}%
\bibitem [{\citenamefont {Mazin}\ \emph {et~al.}(2008)\citenamefont {Mazin},
  \citenamefont {Singh}, \citenamefont {Johannes},\ and\ \citenamefont
  {Du}}]{MazinPRL2008}%
  \BibitemOpen
  \bibfield  {author} {\bibinfo {author} {I.~I. Mazin}, \bibinfo {author}
  {D.~J. Singh}, \bibinfo {author} {M.~D. Johannes}\ and\ \bibinfo {author}
  {M.~H. Du},\ }\bibfield  {title} {\bibinfo {title} {Unconventional
  superconductivity with a sign reversal in the order parameter of
  {LaFeAsO}$_{1-x}${F}$_{x}$},\ }\href
  {https://doi.org/10.1103/PhysRevLett.101.057003} {\bibfield  {journal}
  {\bibinfo  {journal} {Phys. Rev. Lett.}\ }\textbf {\bibinfo {volume} {101}},\
  \bibinfo {pages} {057003} (\bibinfo {year} {2008})}\BibitemShut {NoStop}%
\bibitem [{\citenamefont {Hirschfeld}\ \emph {et~al.}(2011)\citenamefont
  {Hirschfeld}, \citenamefont {Korshunov},\ and\ \citenamefont
  {Mazin}}]{hirschfeld2011}%
  \BibitemOpen
  \bibfield  {author} {\bibinfo {author} {P.~Hirschfeld}, \bibinfo {author}
  {M.~Korshunov}\ and\ \bibinfo {author} {I.~Mazin},\ }\bibfield  {title}
  {\bibinfo {title} {Gap symmetry and structure of fe-based superconductors},\
  }\href@noop {} {\bibfield  {journal} {\bibinfo  {journal} {Reports on
  Progress in Physics}\ }\textbf {\bibinfo {volume} {74}},\ \bibinfo {pages}
  {124508} (\bibinfo {year} {2011})}\BibitemShut {NoStop}%
\bibitem [{\citenamefont {Boeri}\ \emph {et~al.}(2008)\citenamefont {Boeri},
  \citenamefont {Dolgov},\ and\ \citenamefont
  {Golubov}}]{PhysRevLett.101.026403}%
  \BibitemOpen
  \bibfield  {author} {\bibinfo {author} {L.~Boeri}, \bibinfo {author} {O.~V.
  Dolgov}\ and\ \bibinfo {author} {A.~A. Golubov},\ }\bibfield  {title}
  {\bibinfo {title} {Is ${\mathrm{lafeaso}}_{1\ensuremath{-}x}{\mathrm{f}}_{x}$
  an electron-phonon superconductor?},\ }\href
  {https://doi.org/10.1103/PhysRevLett.101.026403} {\bibfield  {journal}
  {\bibinfo  {journal} {Phys. Rev. Lett.}\ }\textbf {\bibinfo {volume} {101}},\
  \bibinfo {pages} {026403} (\bibinfo {year} {2008})}\BibitemShut {NoStop}%
\bibitem [{\citenamefont {Nunner}\ \emph {et~al.}(1999)\citenamefont {Nunner},
  \citenamefont {Schmalian},\ and\ \citenamefont {Bennemann}}]{ybco_nunner}%
  \BibitemOpen
  \bibfield  {author} {\bibinfo {author} {T.~S. Nunner}, \bibinfo {author}
  {J.~Schmalian}\ and\ \bibinfo {author} {K.~H. Bennemann},\ }\bibfield
  {title} {\bibinfo {title} {Influence of electron-phonon interaction on
  spin-fluctuation-induced superconductivity},\ }\href
  {https://doi.org/10.1103/PhysRevB.59.8859} {\bibfield  {journal} {\bibinfo
  {journal} {Phys. Rev. B}\ }\textbf {\bibinfo {volume} {59}},\ \bibinfo
  {pages} {8859} (\bibinfo {year} {1999})}\BibitemShut {NoStop}%
\bibitem [{\citenamefont {Ortenzi}\ \emph {et~al.}(2011)\citenamefont
  {Ortenzi}, \citenamefont {Biermann}, \citenamefont {Andersen}, \citenamefont
  {Mazin},\ and\ \citenamefont {Boeri}}]{cubiso_boeri}%
  \BibitemOpen
  \bibfield  {author} {\bibinfo {author} {L.~Ortenzi}, \bibinfo {author}
  {S.~Biermann}, \bibinfo {author} {O.~K. Andersen}, \bibinfo {author} {I.~I.
  Mazin}\ and\ \bibinfo {author} {L.~Boeri},\ }\bibfield  {title} {\bibinfo
  {title} {Competition between electron-phonon coupling and spin fluctuations
  in superconducting hole-doped cubiso},\ }\href
  {https://doi.org/10.1103/PhysRevB.83.100505} {\bibfield  {journal} {\bibinfo
  {journal} {Phys. Rev. B}\ }\textbf {\bibinfo {volume} {83}},\ \bibinfo
  {pages} {100505} (\bibinfo {year} {2011})}\BibitemShut {NoStop}%
\bibitem [{\citenamefont {Dolgov}\ \emph {et~al.}(2005)\citenamefont {Dolgov},
  \citenamefont {Mazin}, \citenamefont {Golubov}, \citenamefont {Savrasov},\
  and\ \citenamefont {Maksimov}}]{dolgov_mgcni3}%
  \BibitemOpen
  \bibfield  {author} {\bibinfo {author} {O.~V. Dolgov}, \bibinfo {author}
  {I.~I. Mazin}, \bibinfo {author} {A.~A. Golubov}, \bibinfo {author} {S.~Y.
  Savrasov}\ and\ \bibinfo {author} {E.~G. Maksimov},\ }\bibfield  {title}
  {\bibinfo {title} {Critical temperature and enhanced isotope effect in the
  presence of paramagnons in phonon-mediated superconductors},\ }\href
  {https://doi.org/10.1103/PhysRevLett.95.257003} {\bibfield  {journal}
  {\bibinfo  {journal} {Phys. Rev. Lett.}\ }\textbf {\bibinfo {volume} {95}},\
  \bibinfo {pages} {257003} (\bibinfo {year} {2005})}\BibitemShut {NoStop}%
\bibitem [{\citenamefont {Witt}\ \emph {et~al.}(2022)\citenamefont {Witt},
  \citenamefont {Pizarro}, \citenamefont {Berges}, \citenamefont {Nomoto},
  \citenamefont {Arita},\ and\ \citenamefont {Wehling}}]{moiresuperlattice}%
  \BibitemOpen
  \bibfield  {author} {\bibinfo {author} {N.~Witt}, \bibinfo {author} {J.~M.
  Pizarro}, \bibinfo {author} {J.~Berges}, \bibinfo {author} {T.~Nomoto},
  \bibinfo {author} {R.~Arita}\ and\ \bibinfo {author} {T.~O. Wehling},\
  }\bibfield  {title} {\bibinfo {title} {Doping fingerprints of spin and
  lattice fluctuations in moir\'e superlattice systems},\ }\href
  {https://doi.org/10.1103/PhysRevB.105.L241109} {\bibfield  {journal}
  {\bibinfo  {journal} {Phys. Rev. B}\ }\textbf {\bibinfo {volume} {105}},\
  \bibinfo {pages} {L241109} (\bibinfo {year} {2022})}\BibitemShut {NoStop}%
\bibitem [{\citenamefont {Pei}\ \emph {et~al.}(2021)\citenamefont {Pei},
  \citenamefont {Yang}, \citenamefont {Gong}, \citenamefont {Wang},
  \citenamefont {Zhao}, \citenamefont {Gao}, \citenamefont {Chen},
  \citenamefont {Yin}, 31itenamefont {Tian}, \citenamefont {Li}, \citenamefont
  {Cao}, \citenamefont {Lei}, \citenamefont {Cheng},\ and\ \citenamefont
  {Qi}}]{pei2021pressureinduced}%
  \BibitemOpen
  \bibfield  {author} {\bibinfo {author} {C.~Pei}, \bibinfo {author} {P.~Yang},
  \bibinfo {author} {C.~Gong}, \bibinfo {author} {Q.~Wang}, \bibinfo {author}
  {Y.~Zhao}, \bibinfo {author} {L.~Gao}, \bibinfo {author} {K.~Chen}, \bibinfo
  {author} {Q.~Yin}, \bibinfo {author} {S.~Tian} et~al.,\ }\bibfield  {title}
  {\bibinfo {title} {Pressure-induced superconductivity in itinerant
  antiferromagnet {CrB}$_2$},\ }\bibfield  {journal} {\bibinfo  {journal}
  {arXiv:2109.15213}\ }\href {https://doi.org/10.48550/ARXIV.2109.15213}
  {10.48550/ARXIV.2109.15213} (\bibinfo {year} {2021})\BibitemShut {NoStop}%
\bibitem [{\citenamefont {Nagamatsu}\ \emph {et~al.}(2001)\citenamefont
  {Nagamatsu}, \citenamefont {Nakagawa}, \citenamefont {Muranaka},
  \citenamefont {Zenitani},\ and\ \citenamefont {Akimitsu}}]{nagamatsu2001}%
  \BibitemOpen
  \bibfield  {author} {\bibinfo {author} {J.~Nagamatsu}, \bibinfo {author}
  {N.~Nakagawa}, \bibinfo {author} {T.~Muranaka}, \bibinfo {author}
  {Y.~Zenitani}\ and\ \bibinfo {author} {J.~Akimitsu},\ }\bibfield  {title}
  {\bibinfo {title} {Superconductivity at 39 k in magnesium diboride},\
  }\href@noop {} {\bibfield  {journal} {\bibinfo  {journal} {nature}\ }\textbf
  {\bibinfo {volume} {410}},\ \bibinfo {pages} {63} (\bibinfo {year}
  {2001})}\BibitemShut {NoStop}%
\bibitem [{\citenamefont {Mazin}\ and\ \citenamefont
  {Antropov}(2003)}]{Mazin_mgb2}%
  \BibitemOpen
  \bibfield  {author} {\bibinfo {author} {I.~Mazin}\ and\ \bibinfo {author}
  {V.~Antropov},\ }\bibfield  {title} {\bibinfo {title} {Electronic structure,
  electron–phonon coupling, and multiband effects in mgb2},\ }\href
  {https://doi.org/https://doi.org/10.1016/S0921-4534(02)02299-2} {\bibfield
  {journal} {\bibinfo  {journal} {Physica C: Superconductivity}\ }\textbf
  {\bibinfo {volume} {385}},\ \bibinfo {pages} {49} (\bibinfo {year}
  {2003})}\BibitemShut {NoStop}%
\bibitem [{\citenamefont {Graser}\ \emph {et~al.}(2009)\citenamefont {Graser},
  \citenamefont {Maier}, \citenamefont {Hirschfeld},\ and\ \citenamefont
  {Scalapino}}]{Graser2009}%
  \BibitemOpen
  \bibfield  {author} {\bibinfo {author} {S.~Graser}, \bibinfo {author} {T.~A.
  Maier}, \bibinfo {author} {P.~J. Hirschfeld}\ and\ \bibinfo {author} {D.~J.
  Scalapino},\ }\bibfield  {title} {\bibinfo {title} {Near-degeneracy of
  several pairing channels in multiorbital models for the {Fe} pnictides},\
  }\href {https://doi.org/10.1088/1367-2630/11/2/025016} {\bibfield  {journal}
  {\bibinfo  {journal} {New Journal of Physics}\ }\textbf {\bibinfo {volume}
  {11}},\ \bibinfo {pages} {025016} (\bibinfo {year} {2009})}\BibitemShut
  {NoStop}%
\bibitem [{\citenamefont {Altmeyer}\ \emph {et~al.}(2016)\citenamefont
  {Altmeyer}, \citenamefont {Guterding}, \citenamefont {Hirschfeld},
  \citenamefont {Maier}, \citenamefont {Valent{\'\i}},\ and\ \citenamefont
  {Scalapino}}]{altmeyer2016}%
  \BibitemOpen
  \bibfield  {author} {\bibinfo {author} {M.~Altmeyer}, \bibinfo {author}
  {D.~Guterding}, \bibinfo {author} {P.~Hirschfeld}, \bibinfo {author} {T.~A.
  Maier}, \bibinfo {author} {R.~Valent{\'\i}}\ and\ \bibinfo {author} {D.~J.
  Scalapino},\ }\bibfield  {title} {\bibinfo {title} {Role of vertex
  corrections in the matrix formulation of the random phase approximation for
  the multiorbital hubbard model},\ }\href@noop {} {\bibfield  {journal}
  {\bibinfo  {journal} {Physical Review B}\ }\textbf {\bibinfo {volume} {94}},\
  \bibinfo {pages} {214515} (\bibinfo {year} {2016})}\BibitemShut {NoStop}%
\bibitem [{\citenamefont {Kreisel}\ \emph {et~al.}(2013)\citenamefont
  {Kreisel}, \citenamefont {Wang}, \citenamefont {Maier}, \citenamefont
  {Hirschfeld},\ and\ \citenamefont {Scalapino}}]{Kreisel2013}%
  \BibitemOpen
  \bibfield  {author} {\bibinfo {author} {A.~Kreisel}, \bibinfo {author}
  {Y.~Wang}, \bibinfo {author} {T.~A. Maier}, \bibinfo {author} {P.~J.
  Hirschfeld}\ and\ \bibinfo {author} {D.~J. Scalapino},\ }\bibfield  {title}
  {\bibinfo {title} {Spin fluctuations and superconductivity in
  {K}${}_{x}${Fe}${}_{2\ensuremath{-}y}${Se}${}_{2}$},\ }\href
  {https://doi.org/10.1103/PhysRevB.88.094522} {\bibfield  {journal} {\bibinfo
  {journal} {Phys. Rev. B}\ }\textbf {\bibinfo {volume} {88}},\ \bibinfo
  {pages} {094522} (\bibinfo {year} {2013})}\BibitemShut {NoStop}%
\bibitem [{\citenamefont {D\"urrnagel}\ \emph {et~al.}(2022)\citenamefont
  {D\"urrnagel}, \citenamefont {Beyer}, \citenamefont {Thomale},\ and\
  \citenamefont {Schwemmer}}]{Durrnagel2022}%
  \BibitemOpen
  \bibfield  {author} {\bibinfo {author} {M.~D\"urrnagel}, \bibinfo {author}
  {J.~Beyer}, \bibinfo {author} {R.~Thomale}\ and\ \bibinfo {author}
  {T.~Schwemmer},\ }\bibfield  {title} {\bibinfo {title} {Unconventional
  superconductivity from weak coupling},\ }\href
  {https://doi.org/10.1140/epjb/s10051-022-00371-4} {\bibfield  {journal}
  {\bibinfo  {journal} {The European Physical Journal B}\ }\textbf {\bibinfo
  {volume} {95}},\ \bibinfo {pages} {112} (\bibinfo {year} {2022})}\BibitemShut
  {NoStop}%
\bibitem [{\citenamefont {Biswas}\ \emph {et~al.}()\citenamefont {Biswas},
  \citenamefont {Kreisel}, \citenamefont {Valadkhani}, \citenamefont
  {D\"urrnagel}, \citenamefont {Schwemmer}, \citenamefont {Thomale},
  \citenamefont {Valent\'i},\ and\ \citenamefont {Mazin}}]{supplemental}%
  \BibitemOpen
  \bibfield  {author} {\bibinfo {author} {S.~Biswas}, \bibinfo {author}
  {A.~Kreisel}, \bibinfo {author} {A.~Valadkhani}, \bibinfo {author}
  {M.~D\"urrnagel}, \bibinfo {author} {T.~Schwemmer}, \bibinfo {author}
  {R.~Thomale}, \bibinfo {author} {R.~Valent\'i}\ and\ \bibinfo {author}
  {I.~Mazin},\ }\bibfield  {title} {\bibinfo {title} {Supplementary material
  for: Hybrid $s$-wave superconductivity in {CrB}$_2$ which gives details of
  the {EPC} and {SF} calculations and contains refs. [xx-yy].},\ }\href@noop {}
  {\ }\BibitemShut {NoStop}%
\bibitem [{\citenamefont {Giannozzi}\ \emph {et~al.}(2009)\citenamefont
  {Giannozzi}, \citenamefont {Baroni}, \citenamefont {Bonini}, \citenamefont
  {Calandra}, \citenamefont {Car}, \citenamefont {Cavazzoni}, \citenamefont
  {Ceresoli}, \citenamefont {Chiarotti}, \citenamefont {Cococcioni},
  \citenamefont {Dabo}, \citenamefont {Corso}, \citenamefont {de~Gironcoli},
  \citenamefont {Fabris}, \citenamefont {Fratesi}, \citenamefont {Gebauer},
  \citenamefont {Gerstmann}, \citenamefont {Gougoussis}, \citenamefont
  {Kokalj}, \citenamefont {Lazzeri}, \citenamefont {Martin-Samos},
  \citenamefont {Marzari}, \citenamefont {Mauri}, \citenamefont {Mazzarello},
  \citenamefont {Paolini}, \citenamefont {Pasquarello}, \citenamefont
  {Paulatto}, \citenamefont {Sbraccia}, \citenamefont {Scandolo}, \citenamefont
  {Sclauzero}, \citenamefont {Seitsonen}, \citenamefont {Smogunov},
  \citenamefont {Umari},\ and\ \citenamefont {Wentzcovitch}}]{QE}%
  \BibitemOpen
  \bibfield  {author} {\bibinfo {author} {P.~Giannozzi}, \bibinfo {author}
  {S.~Baroni}, \bibinfo {author} {N.~Bonini}, \bibinfo {author} {M.~Calandra},
  \bibinfo {author} {R.~Car}, \bibinfo {author} {C.~Cavazzoni}, \bibinfo
  {author} {D.~Ceresoli}, \bibinfo {author} {G.~L. Chiarotti}, \bibinfo
  {author} {M.~Cococcioni} et~al.,\ }\bibfield  {title} {\bibinfo {title}
  {{QUANTUM ESPRESSO}: a modular and open-source software project for quantum
  simulations of materials},\ }\href
  {https://doi.org/https://doi.org/10.1088/0953-8984/21/39/395502} {\bibfield
  {journal} {\bibinfo  {journal} {Journal of Physics: Condensed Matter}\
  }\textbf {\bibinfo {volume} {21}},\ \bibinfo {pages} {395502} (\bibinfo
  {year} {2009})}\BibitemShut {NoStop}%
\bibitem [{\citenamefont {Kortus}\ \emph {et~al.}(2001)\citenamefont {Kortus},
  \citenamefont {Mazin}, \citenamefont {Belashchenko}, \citenamefont
  {Antropov},\ and\ \citenamefont {Boyer}}]{kortus2001}%
  \BibitemOpen
  \bibfield  {author} {\bibinfo {author} {J.~Kortus}, \bibinfo {author}
  {I.~Mazin}, \bibinfo {author} {K.~D. Belashchenko}, \bibinfo {author} {V.~P.
  Antropov}\ and\ \bibinfo {author} {L.~Boyer},\ }\bibfield  {title} {\bibinfo
  {title} {Superconductivity of metallic boron in mgb 2},\ }\href@noop {}
  {\bibfield  {journal} {\bibinfo  {journal} {Physical Review Letters}\
  }\textbf {\bibinfo {volume} {86}},\ \bibinfo {pages} {4656} (\bibinfo {year}
  {2001})}\BibitemShut {NoStop}%
\bibitem [{\citenamefont {Bauer}\ \emph {et~al.}(2014)\citenamefont {Bauer},
  \citenamefont {Regnat}, \citenamefont {Blum}, \citenamefont
  {Gottlieb-Sch\"onmeyer}, \citenamefont {Pedersen}, \citenamefont {Meven},
  \citenamefont {Wurmehl}, \citenamefont {Kune\ifmmode~\check{s}\else
  \v{s}\fi{}},\ and\ \citenamefont {Pfleiderer}}]{bauer}%
  \BibitemOpen
  \bibfield  {author} {\bibinfo {author} {A.~Bauer}, \bibinfo {author}
  {A.~Regnat}, \bibinfo {author} {C.~G.~F. Blum}, \bibinfo {author}
  {S.~Gottlieb-Sch\"onmeyer}, \bibinfo {author} {B.~Pedersen}, \bibinfo
  {author} {M.~Meven}, \bibinfo {author} {S.~Wurmehl}, \bibinfo {author}
  {J.~Kune\ifmmode~\check{s}\else \v{s}\fi{}}\ and\ \bibinfo {author}
  {C.~Pfleiderer},\ }\bibfield  {title} {\bibinfo {title} {Low-temperature
  properties of single-crystal ${\mathrm{crb}}_{2}$},\ }\href
  {https://doi.org/10.1103/PhysRevB.90.064414} {\bibfield  {journal} {\bibinfo
  {journal} {Phys. Rev. B}\ }\textbf {\bibinfo {volume} {90}},\ \bibinfo
  {pages} {064414} (\bibinfo {year} {2014})}\BibitemShut {NoStop}%
\bibitem [{\citenamefont {Basov}\ and\ \citenamefont
  {Chubukov}(2011)}]{Basov2011}%
  \BibitemOpen
  \bibfield  {author} {\bibinfo {author} {D.~N. Basov}\ and\ \bibinfo {author}
  {A.~V. Chubukov},\ }\bibfield  {title} {\bibinfo {title} {Manifesto for a
  higher ${T}_c$},\ }\href {https://doi.org/10.1038/nphys1975} {\bibfield
  {journal} {\bibinfo  {journal} {Nature Physics}\ }\textbf {\bibinfo {volume}
  {7}},\ \bibinfo {pages} {272} (\bibinfo {year} {2011})}\BibitemShut {NoStop}%
\bibitem [{\citenamefont {Koepernik}\ and\ \citenamefont
  {Eschrig}(1999)}]{koepernik_1999}%
  \BibitemOpen
  \bibfield  {author} {\bibinfo {author} {K.~Koepernik}\ and\ \bibinfo {author}
  {H.~Eschrig},\ }\bibfield  {title} {\bibinfo {title} {Full-potential
  nonorthogonal local-orbital minimum-basis band-structure scheme},\ }\href
  {https://doi.org/10.1103/PhysRevB.59.1743} {\bibfield  {journal} {\bibinfo
  {journal} {Phys. Rev. B}\ }\textbf {\bibinfo {volume} {59}},\ \bibinfo
  {pages} {1743} (\bibinfo {year} {1999})}\BibitemShut {NoStop}%
\bibitem [{\citenamefont {Perdew}\ and\ \citenamefont
  {Wang}(1992)}]{perdew1992}%
  \BibitemOpen
  \bibfield  {author} {\bibinfo {author} {J.~P. Perdew}\ and\ \bibinfo {author}
  {Y.~Wang},\ }\bibfield  {title} {\bibinfo {title} {Accurate and simple
  analytic representation of the electron-gas correlation energy},\ }\href
  {https://doi.org/10.1103/PhysRevB.45.13244} {\bibfield  {journal} {\bibinfo
  {journal} {Phys. Rev. B}\ }\textbf {\bibinfo {volume} {45}},\ \bibinfo
  {pages} {13244} (\bibinfo {year} {1992})}\BibitemShut {NoStop}%
\bibitem [{\citenamefont {Kresse}\ and\ \citenamefont {Hafner}(1993)}]{vasp}%
  \BibitemOpen
  \bibfield  {author} {\bibinfo {author} {G.~Kresse}\ and\ \bibinfo {author}
  {J.~Hafner},\ }\bibfield  {title} {\bibinfo {title} {Ab initio molecular
  dynamics for liquid metals},\ }\href@noop {} {\bibfield  {journal} {\bibinfo
  {journal} {Phys. Rev. B}\ }\textbf {\bibinfo {volume} {47}},\ \bibinfo
  {pages} {558} (\bibinfo {year} {1993})}\BibitemShut {NoStop}%
\bibitem [{\citenamefont {Baroni}\ \emph {et~al.}(2001)\citenamefont {Baroni},
  \citenamefont {de~Gironcoli}, \citenamefont {Dal~Corso},\ and\ \citenamefont
  {Giannozzi}}]{baroniDFPT}%
  \BibitemOpen
  \bibfield  {author} {\bibinfo {author} {S.~Baroni}, \bibinfo {author}
  {S.~de~Gironcoli}, \bibinfo {author} {A.~Dal~Corso}\ and\ \bibinfo {author}
  {P.~Giannozzi},\ }\bibfield  {title} {\bibinfo {title} {Phonons and related
  crystal properties from density-functional perturbation theory},\ }\href
  {https://doi.org/10.1103/RevModPhys.73.515} {\bibfield  {journal} {\bibinfo
  {journal} {Rev. Mod. Phys.}\ }\textbf {\bibinfo {volume} {73}},\ \bibinfo
  {pages} {515} (\bibinfo {year} {2001})}\BibitemShut {NoStop}%
 \bibitem [{\citenamefont {McMillan}(1968)}]{McMillan}%
   \BibitemOpen
   \bibfield  {author} {\bibinfo {author} {W.~L. McMillan},\ }\bibfield  {title}
   {\bibinfo {title} {Transition temperature of strong-coupled
   superconductors},\ }\href {https://doi.org/10.1103/PhysRev.167.331}
   {\bibfield  {journal} {\bibinfo  {journal} {Phys. Rev.}\ }\textbf {\bibinfo
   {volume} {167}},\ \bibinfo {pages} {331} (\bibinfo {year}
   {1968})}\BibitemShut {NoStop}%
 \bibitem [{\citenamefont {Allen}\ and\ \citenamefont
   {Dynes}(1975)}]{Allen-Dynes}%
   \BibitemOpen
   \bibfield  {author} {\bibinfo {author} {P.~B. Allen}\ and\ \bibinfo {author}
   {R.~C. Dynes},\ }\bibfield  {title} {\bibinfo {title} {Transition temperature
   of strong-coupled superconductors reanalyzed},\ }\href
   {https://doi.org/10.1103/PhysRevB.12.905} {\bibfield  {journal} {\bibinfo
   {journal} {Phys. Rev. B}\ }\textbf {\bibinfo {volume} {12}},\ \bibinfo
   {pages} {905} (\bibinfo {year} {1975})}\BibitemShut {NoStop}%
 \bibitem [{\citenamefont {Beyer}\ \emph {et~al.}(2022)\citenamefont {Beyer},
   \citenamefont {Hauck},\ and\ \citenamefont {Klebl}}]{Beyer2022}%
   \BibitemOpen
   \bibfield  {author} {\bibinfo {author} {J.~Beyer}, \bibinfo {author} {J.~B.
   Hauck}\ and\ \bibinfo {author} {L.~Klebl},\ }\bibfield  {title} {\bibinfo
   {title} {Reference results for the momentum space functional renormalization
   group},\ }\href {https://doi.org/10.1140/epjb/s10051-022-00323-y} {\bibfield
   {journal} {\bibinfo  {journal} {The European Physical Journal B}\ }\textbf
   {\bibinfo {volume} {95}},\ \bibinfo {pages} {65} (\bibinfo {year}
   {2022})}\BibitemShut {NoStop}%
\end{thebibliography}%
%\nocite{koepernik_1999,perdew1992,Graser2009,Kreisel2013, Durrnagel2022,vasp,baroniDFPT}

% Code to add supplement file at the end of the document
%\ifarXiv
%    \foreach \x in {1,...,\numbersupplementpages}
%    {
%        \clearpage
%        \includepdf[pages={\x,{}}]{\supplementfilename}
%    }
%\fi

\clearpage

%\clearpage
\widetext
%\appendix
\begin{center}
\textbf{Supplemental Material: Hybrid $s$-wave superconductivity in \crb }
 \bigskip
\end{center}
\onecolumngrid
\setcounter{page}{1}
\makeatletter

\setlength{\textheight}{9.5in}
        \setcounter{table}{0}
        \renewcommand{\thetable}{S\arabic{table}}%
        \setcounter{figure}{0}
        \renewcommand{\thefigure}{S\arabic{figure}}%

%\title{Supplemental Material: Hybrid $s$-wave superconductivity in \crb }
\author{Sananda Biswas}
%\email{biswas@itp.uni-frankfurt.de}
%%\authornote{Corresponding authors}
%\affiliation{Institut f\"ur Theoretische Physik, Goethe-Universit\"at Frankfurt, 60438 Frankfurt am Main, Germany}
%\author{Andreas Kreisel}
%\email{kreisel@itp.uni-leipzig.de}
%%\authornote{Corresponding authors}
%\affiliation{Institut  f\"ur Theoretische Physik, Universit\"at Leipzig, Br\"uderstr. 16, 04103 Leipzig, Germany}
%\affiliation{Niels Bohr Institute, University of Copenhagen, 2100 Copenhagen, Denmark}
%\author{Adrian Valadkhani}
%\affiliation{Institut f\"ur Theoretische Physik, Goethe-Universit\"at Frankfurt, 60438 Frankfurt am Main, Germany}
%\author{Matteo D\"urrnagel}
%%\affiliation{Julius-Maximilians-Universit\"at W\"urzburg, W\"urzburg, Germany}
%\author{Tilman Schwemmer}
%%\affiliation{Julius-Maximilians-Universit\"at W\"urzburg, W\"urzburg, Germany}
%\author{Ronny Thomale}
%\affiliation{Julius-Maximilians-Universit\"at W\"urzburg, W\"urzburg, Germany}
%\author{Roser Valent\'i}
%\affiliation{Institut f\"ur Theoretische Physik, Goethe-Universit\"at Frankfurt, 60438 Frankfurt am Main, Germany}
%\author{Igor I. Mazin}
%\affiliation{Department of Physics and Astronomy, George Mason University, Fairfax, VA 22030}
%\affiliation{Quantum Science and Engineering Center, George Mason University, Fairfax, VA 22030}

%\begin{abstract} 
{\small This Supplementary Material contains details on the calculations for the spin-fluctuation pairing and electron phonon paring and discusses the charge density wave states. We include a section on a simplified two band model capable to capture the hybrid pairing properties as spin-fluctuations and electron phonon pairing gets modified by pressure and discuss feasibility of a functional renormalization group approach for the present system.}
%\end{abstract}
%\date{\today}
%\maketitle

%

\section{Band structure as function of pressure}

\begin{figure}[!b]
\includegraphics[width=0.5\linewidth]{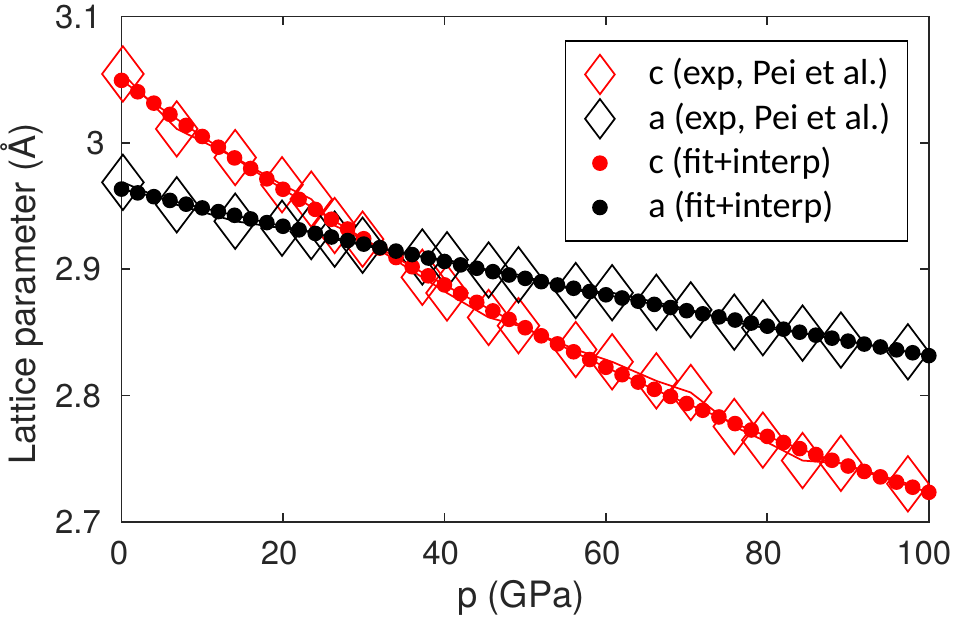}
\caption{Lattice parameters $c$ and $a$ in \AA{} as used for the DFT calculation to obtain the tight binding model. The open diamonds are experimental data and full circles show the fit using a polynomial of second order and interpolation to obtain lattice constants at increments of 2 GPa. %\SB{As you now have interpolated lattice parameters in the interval of 2 GPa, shall we update this figure?}
}
\label{fig_exp}
\end{figure}

 \begin{figure}[!t]
\includegraphics[width=\linewidth]{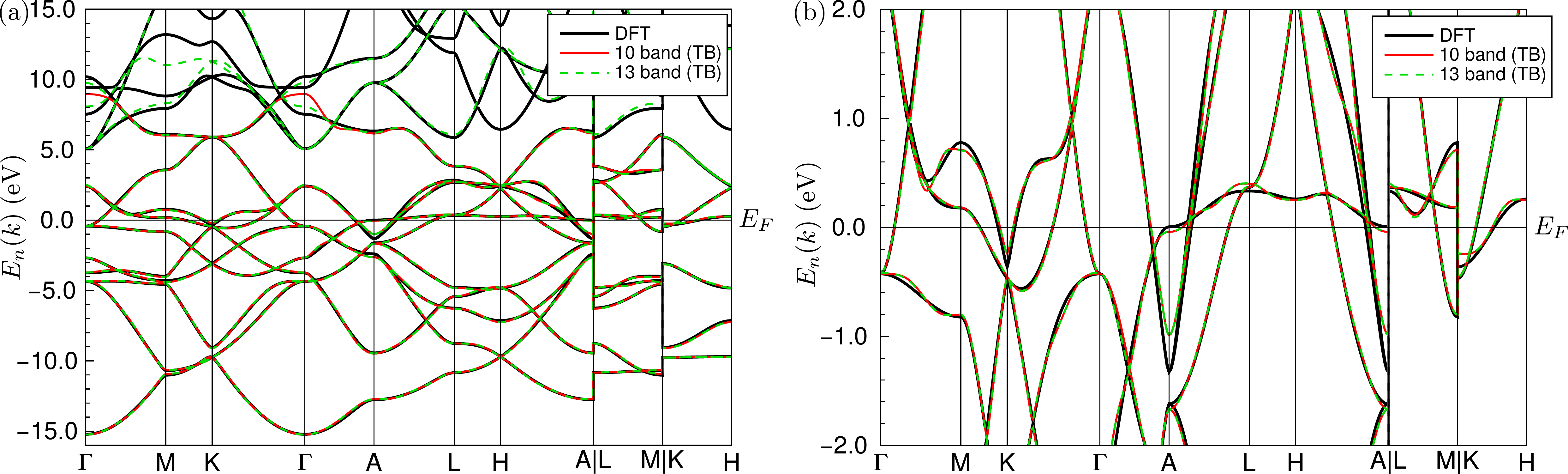}
\caption{Wannier downfolding. Comparison of DFT derived bands (black) with a downfolding to 10 bands (red) and downfolding to 13 bands (green, dashed) where also the antibonding orbitals are taken into account. (a) Bands at large energy scale and (b) blowup close to the Fermi level such that small deviations can be observed for few bands. (Calculation for a pressure of 25 GPa.)\label{fig_bands_compare}}
\end{figure}

The starting point for the spin-fluctuation pairing calculations is a tight binding representation of the electronic structure as a function of pressure which we obtain as follows. Starting from the lattice constants as determined experimentally in \cite{pei2021pressureinduced}, we fit these with a polynomial of second order to obtain a smooth behavior of the lattice constants as a function of pressure in order to avoid non-monotonous evolution of the Fermi surface and low energy band structure. In Fig. \ref{fig_exp}, we show the experimental data together with the lattice constants as used in our {\it ab initio} calculations using the full-potential local-orbital (FPLO) code - version 18.00-52 ~\cite{koepernik_1999} with the non-relativistic LSDA ("Perdew Wang 92") approximation~\cite{perdew1992}. The crystal structure of  \crb\ belongs to the space group $P6/mmm$ (\# 191) and the Wyckoff positions of Cr and B atoms are $(0,0,0)$ and $(1/3, 2/3, 1/2)$, respectively. A $\mathbf{k}$ -grid of $12\times12\times12$ is used. Convergence is checked with respect to the $\mathbf{k}$-grid  and relativistic effects are found to have negligible influence on the band structure close to the Fermi level.

For the downfolding to a tight binding model, we use the following initial projections of all five Cr 3$d$ orbitals, the two B $p_z$ orbitals and three bonding $sp_2$ orbitals of the two B atoms (10 orbital model). For a 13 orbital model, we additionally take into account three antibonding $sp_2$ orbitals. This downfolding yields a tight binding model with roughly 20000 real-valued hoppings (for the 10 band model) up to imaginary parts of relative order $10^{-8}$ which are set to zero. This  Hamiltonian can be written in momentum space as,
\begin{equation}
H_0=\sum_{\mathbf{k}\sigma \ell \ell'} t^{\ell \ell'}_{\mathbf{k}} c_{\ell\sigma}^\dagger(\mathbf{k}) c_{\ell'\sigma}(\mathbf{k}),
 \label{eq_tb}
\end{equation}
where $c_{\ell\sigma}^\dagger(\mathbf{k})$ is the Fourier amplitude of an operator $c_{i\ell\sigma}^\dagger$ that creates an electron in Wannier orbital $\ell$ with spin $\sigma$ and $t^{\ell \ell'}_{\mathbf{k}}$ is the Fourier transform of the hopping elements connecting states $\ell$ and $\ell'$. The dependence on the cutoff in distance and energy has been checked and set to \verb|ham_cutoff 20.0| and \verb|WF_ham_threshold 0.0001| to yield the tight binding bands as shown in Fig. \ref{fig_bands_compare}.

\section{Spin-fluctuation calculations}
\begin{figure*}[tb]
 \includegraphics[width=0.8\linewidth]{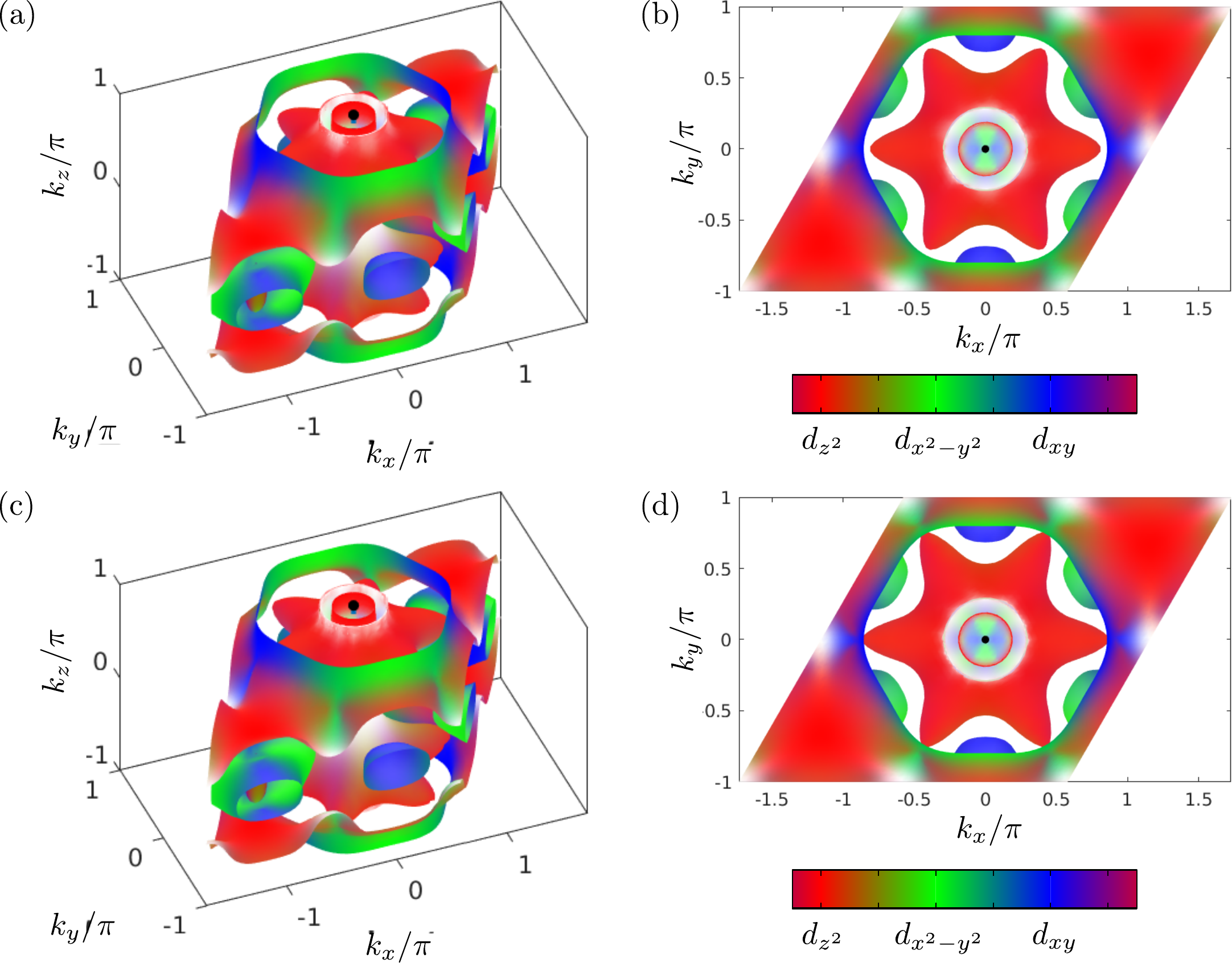}
\caption{Fermi surface from the 10 band model (a-b) and the 13 band model (c-d) for $p=25\,\text{GPa}$. Colors represent the different Wannier states as indicated in the legend, bright color indicates significant weight of other states.\label{fig_fs_compare}}
\end{figure*}

\begin{figure*}[tb]
\includegraphics[width=\textwidth]{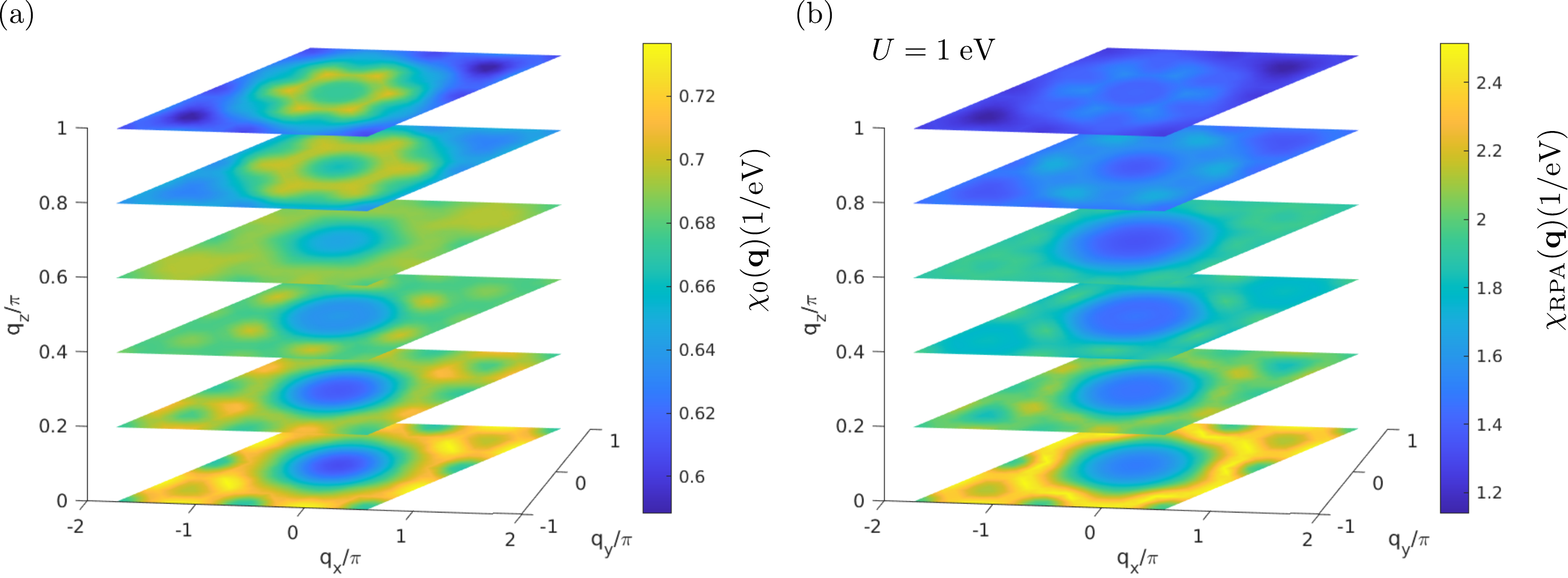}
\caption{Bare susceptibility calculated for $p=24 \;\text{GPa}$ (a) and RPA susceptibility (b) in the Brillouin-zone showing that the RPA modifies the peak structure.}
\end{figure*}

% \begin{figure}[!tbp]
% \includegraphics[width=0.33\textwidth]{}
% \includegraphics[width=0.33\textwidth]{}
% \caption{Template_figure}
% \end{figure}

\begin{figure*}[tb]
\includegraphics[width=\textwidth]{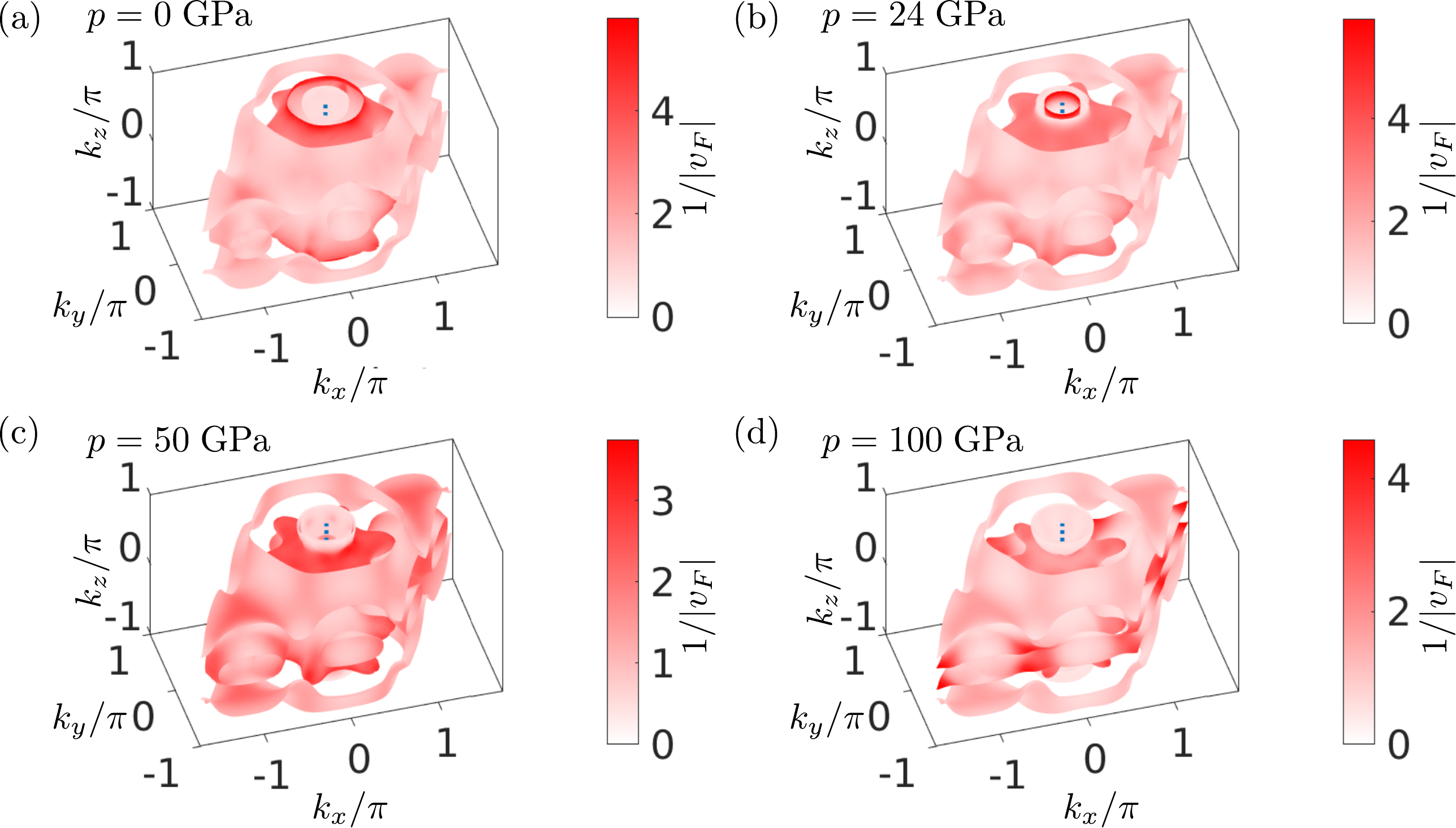}
\caption{Examples of Fermi surfaces with inverse Fermi velocity for pressures 0, 24, 50, 100 GPa.}
\end{figure*}

The spin-fluctuation pairing calculations are based on the tight-binding formulation, as given in Eq.~(\ref{eq_tb}), where we start to calculate the two-point functions, i.e. the (generalized) susceptibility in the paramagnetic state~\cite{Graser2009},
\begin{align}
	\chi_{\ell_1 \ell_2 \ell_3 \ell_4}^0 (q) & = - \sum_{k,\mu,\nu } M_{\ell_1 \ell_2 \ell_3 \ell_4}^{\mu\nu} (\mathbf{k},\mathbf{q})  G^{\mu} (k+q) G^{\nu} (k)\,.   \label{eqn_supersuscept}
\end{align}
Here, we have adopted the shorthand notation $k\equiv (\mathbf{k},\omega_n)$ for the momentum and frequency. The matrix elements are given by,
\begin{align}
	 M_{\ell_1 \ell_2 \ell_3 \ell_4}^{\mu\nu} (\mathbf{k},\mathbf{q})=
	 %& =&
	 a_\nu^{\ell_4} (\mathbf{k}) a_\nu^{\ell_2,*} (\mathbf{k}) a_\mu^{\ell_1} (\mathbf{k}+\mathbf{q}) a_\mu^{\ell_3,*} (\mathbf{k}+\mathbf{q}),\label{eq_M}
\end{align}
and the Green's function in band space reads as:
\begin{align}
  G^\mu(\mathbf{k},\omega_n)=[{i \omega_n -  E_\mu(\mathbf{k})}]^{-1}.
\end{align}
Next, we calculate the interacting susceptibility in a random phase approximation (RPA), where bubble diagrams are
partially re-summed to get
\begin{align}
\label{eqn:RPA}
 \chi_{(0,1)\,\ell_1\ell_2\ell_3\ell_4}^{\rm RPA} (\mathbf{q},\omega) &= \left\{\chi^0 (\mathbf{q},\omega) \left[1 -\bar U^{(s,c)} \chi^0 (\mathbf{q},\omega) \right]^{-1} \right\}_{\ell_1\ell_2\ell_3\ell_4}.%\\
\end{align}
In this equation, $\bar U^{(s,c)}$ is the matrix for the generalized spin ($s$) and charge ($c$) interactions containing the parameters of the Hubbard-Kanamouri Hamiltonian, $U,U',J,J'$, which we choose to be (spin) rotationally invariant, $U'=U-2J$, $J=J'$ and nonzero only in the Cr $3d$ orbital components.
This choice is guided by the expectation that the B orbitals are less correlated and it avoids possible complications originating from the bonding/antibonding B orbitals as these are not centered at an atomic position. Moreover, a `fully local' interaction Hamiltonian would not be an appropriate starting point. The pairing interaction in the orbital space in given by~\cite{Graser2009, Kreisel2013, Durrnagel2022}, 
\begin{align}\label{eq_fullGammanodress}
 	&{\Gamma}_{\ell_1\ell_2\ell_3\ell_4} (\mathbf{k},\mathbf{k}')\! =\!\frac 12 \Bigl[3\bar U^s \chi_1^{\rm RPA} (\mathbf{k}-\mathbf{k}') \bar U^s %\right.\,\,\notag\\
 	%&\,\left. 
 	+   \bar U^s - \bar U^c \chi_0^{\rm RPA} (\mathbf{k}-\mathbf{k}') \bar U^c +  \bar U^c \Bigr]_{\ell_1\ell_2\ell_3\ell_4},%\nonumber 
\end{align}
where $\chi_0^{\rm RPA}$ is the charge susceptibility and $\chi_1^{\rm RPA}$ the spin susceptibility in RPA approximation \cite{Graser2009} at zero frequency. We deduce the leading and sub-leading superconducting instabilities from solving the linearized gap equation (parametrized on three-dimensional Fermi surfaces \cite{Kreisel2013,Durrnagel2022}),
 \begin{equation}\label{eqn:gapeqn}
-\frac{1}{V_G}  \sum_\mu\int_{\text{FS}_\mu}dS'\; \Gamma_{\nu\mu}(\mathbf{k},\mathbf{k}') \frac{ g_i(\mathbf{k}')}{|v_{\text{F}\mu}(\mathbf{k}')|}=\lambda_i g_{i}(\mathbf{k})\,,
 \end{equation}
 for the eigenvalues $\lambda_i$ and the eigenvectors $g_i(\mathbf{k})$. The pairing interaction is projected into band space by
 \begin{align}
	&{\Gamma}_{\nu\mu} (\mathbf{k},\mathbf{k}')  = \mathrm{Re}\sum_{\ell_1\ell_2\ell_3\ell_4} a_{\nu}^{\ell_1,*}(\mathbf{k}) a_{\nu}^{\ell_4,*}(-\mathbf{k})
	%\nonumber\\	&\hspace{0.6cm}\times 
	{\Gamma}_{\ell_1\ell_2\ell_3\ell_4} (\mathbf{k},\mathbf{k}') \;  a_{\mu}^{\ell_2}(\mathbf{k}') a_{\mu}^{\ell_3}(-\mathbf{k}')\, \label{eq_Gam_mu_nu}\,,
\end{align}
where $a_{\nu}^{\ell}(\mathbf{k})$ is the matrix element of orbital $\nu$ for the unitary transformation to band space of band $\ell$.
The integral is done over the Fermi surface $dS$ with total surface $V_G$ and weights are given by the inverse Fermi velocity $v_{\text{F}\mu}(\mathbf{k})$. The leading instability, identified by the largest eigenvalue $\lambda_i$, will lead to a superconducting order parameter $\Delta(\mathbf{k})$ that is proportional to $g_i(\mathbf{k})$ at $T_c$.

\subsubsection*{Tests on model dependence}
The pairing calculation based on a 13 orbital model using a complex three-dimensional Fermi surface geometry is, although challenging,  numerically feasible. However, for a sweep of the pressure dependence, we restricted our analysis to the 10 orbital tight binding model only and additionally constrained the calculation of the susceptibility, Eq. (\ref{eqn_supersuscept}), to the components of the 5 Cr-$d$ orbitals. This approximation is expected to be very good given that the partial density of states of the other 5 (or 8 for the 13 orbital model) orbitals at low energies is very small and thus the respective components of the susceptibility tensor are small as well. In the pairing vertex (in our approximation), the spin and charge matrices, $\bar U^{(s,c)}$, only have nonzero elements for the correlated Cr-$d$ orbitals. Therefore the only correction from the B-type orbitals are originating from the off-diagonal elements in the RPA approach, i.e. from inversion of the matrix in Eq. (\ref{eqn:RPA}). We checked the validity of the previous arguments by performing  calculations using all components in the 10 and 13 orbital model at a selected pressure and comparing the results for the pairing eigenvalues $\lambda_i$s and pairing states $g_i(\mathbf{k})$s. These values come out to be very similar.

\section{Electron-phonon calculations}
\begin{figure}[!b]
\includegraphics[width=0.49\textwidth]{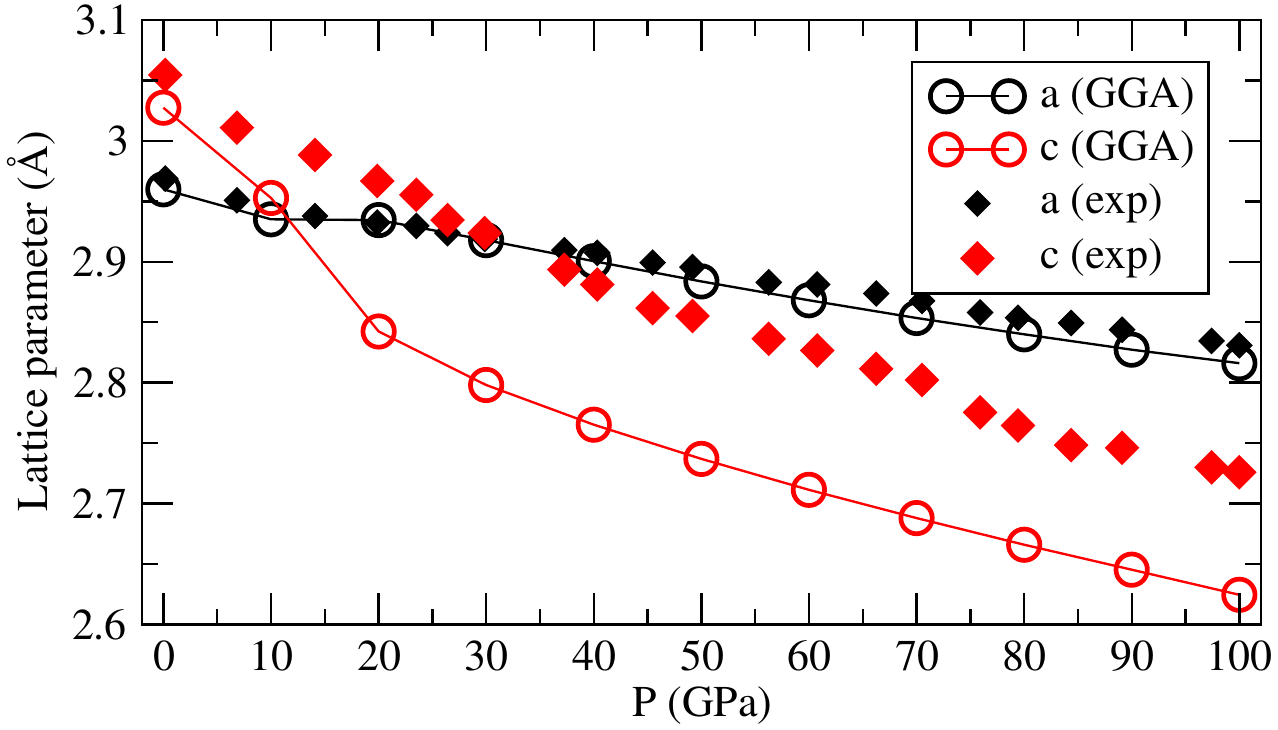}
\caption{Comparison of experimental and calculated lattice parameters. The calculated in-plane lattice parameters, $a=b$, compares very well with experiment. However, the calculated parameters $c$ always remains lower than the experimental value. }
\label{fig:lat_para_cal}
\end{figure}
\begin{figure}[!t]
\includegraphics[width=0.49\textwidth]{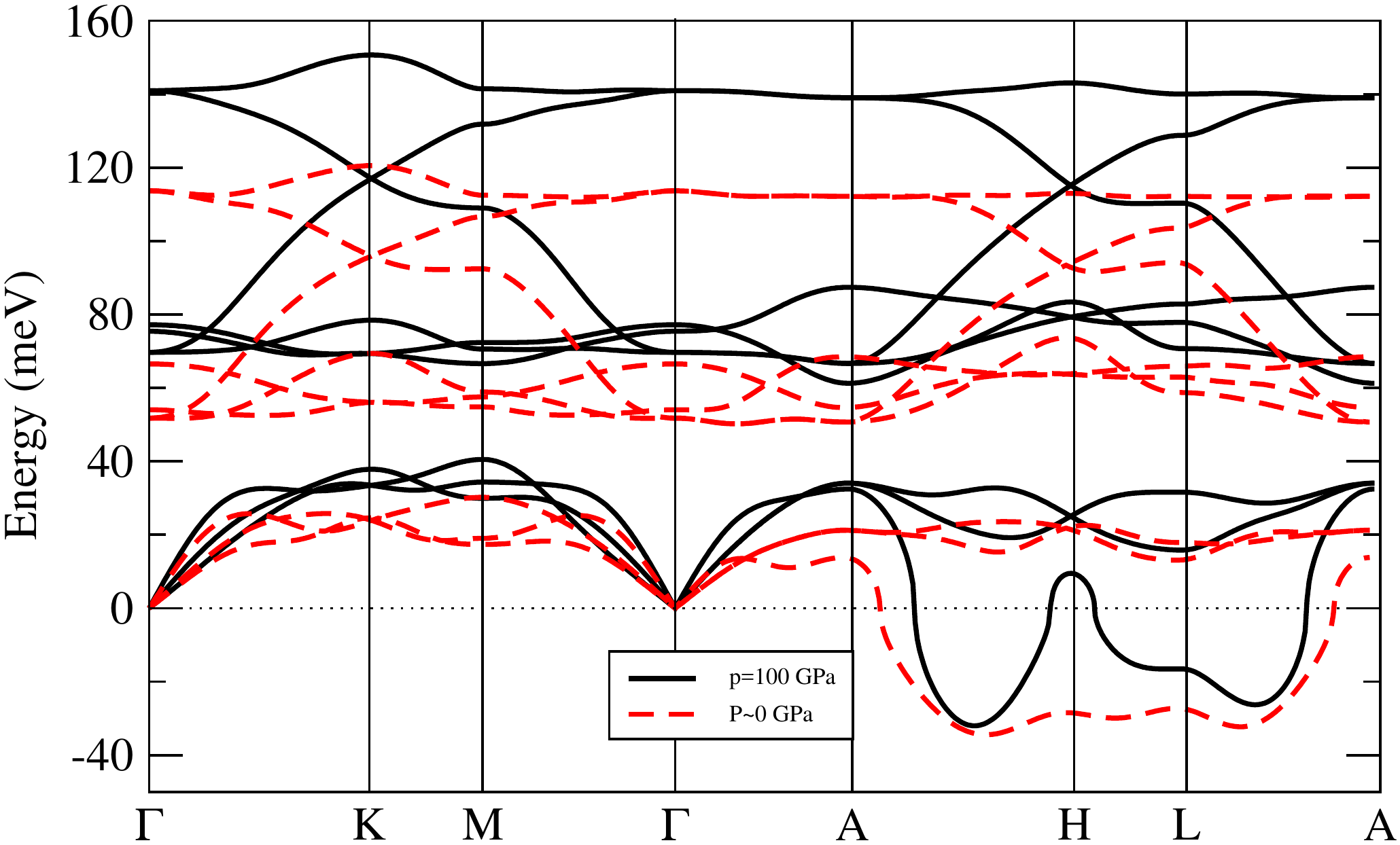}
\caption{Phonon dispersion at ambient (red dashed lines) and 100 GPa (black solid lines) pressure with structures at experimental lattice constants.  }
\label{fig:ph_disp_pressure}
\end{figure}

All the geometric optimizations (relaxations) at fixed pressures are performed using density functional theory (DFT) as implemented in Vienna {\it ab initio} simulation package (vasp)~\cite{vasp} with plane-wave basis set (cutoff of 500 eV) and projected augmented wave (PAW) pseudopotentials. A $12 \times 12 \times12$ Monkhorst-Pack $\mathbf{K}$-grid is employed to perform structural relaxations until the maximum force for each component on each individual atom is smaller than 0.001 eV/\AA. We find that the relaxed structures differ only on the lattice parameter $c$ with respect to the experimental lattice parameters. This explains the reason behind finding a phonon instability of the experimental structure along the crystallographic direction  $c$, as shown below.

We have performed {\it ab initio} density functional perturbation theory (DFPT)~\cite{baroniDFPT} calculations in order to obtain the phonon dispersions and electron-phonon coupling (EPC) constants, as implemented in Quantum ESPRESSO~\cite{QE}. We used plane waves basis sets with cutoff of 80 Ry (and of 800 Ry for the corresponding charge densities) in combination with ultrasoft pseudopotentials in the generalized gradient approximations (GGA).  

Using the above parameters, electron-phonon coefficients are calculated from the derivative of the self-consistent Kohn-Sham potentials, $V_\mathrm{SCF}$, using
\begin{equation}
 g_{mn,\nu}(\mathbf{k,q})= \frac{1}{\sqrt{2\omega_{q,\nu}}}\langle{\Psi_{m,\mathbf{k+q}}} |\partial_{\mathbf{q},\nu}V_{\mathrm{SCF}} |{\Psi_{n,\mathbf{k}}} \rangle,
\end{equation}
\noindent
where $m,n$ are the band indices and $\mathbf{k},\mathbf{q}$ are the electron and phonon wavevectors.

The electron-phonon line-width and the spectral function $\alpha^2F$ are then calculated using the following equations:\\

\begin{align}
{\gamma_{\mathbf{q},\nu}}&=2\pi \omega_{\mathbf{q},\nu}\sum_{m,n}\int \frac{d^3k}{\Omega_{BZ} } |{g_{mn,\nu}(\mathbf{k,q})}|^2 \delta(\epsilon_{m,\mathbf{k+q}}-\epsilon_F)\delta(\epsilon_{n,\mathbf{k}}-\epsilon_F) \mathrm{ and}\\ 
{\alpha^2F(\omega)}&= \frac{1}{2\pi N_F} \sum_{\mathbf{q},\nu} \delta (\omega-\omega_{\mathbf{q},\nu}) \frac{{\gamma_{\mathbf{q},\nu}}}{\hbar \omega_{\mathbf{q},\nu}}.
\end{align}

\noindent
EPC constants are then calculated by, 
\begin{equation}
{\lambda}=2\int_0^{\omega_{\rm max}} {\alpha^2F(\omega)} d\omega, 
\end{equation}
\noindent
where $\omega_{\rm max}$ is the maximum phonon frequency at a given pressure.

\begin{figure*}[!t]
\includegraphics[width=0.94\textwidth]{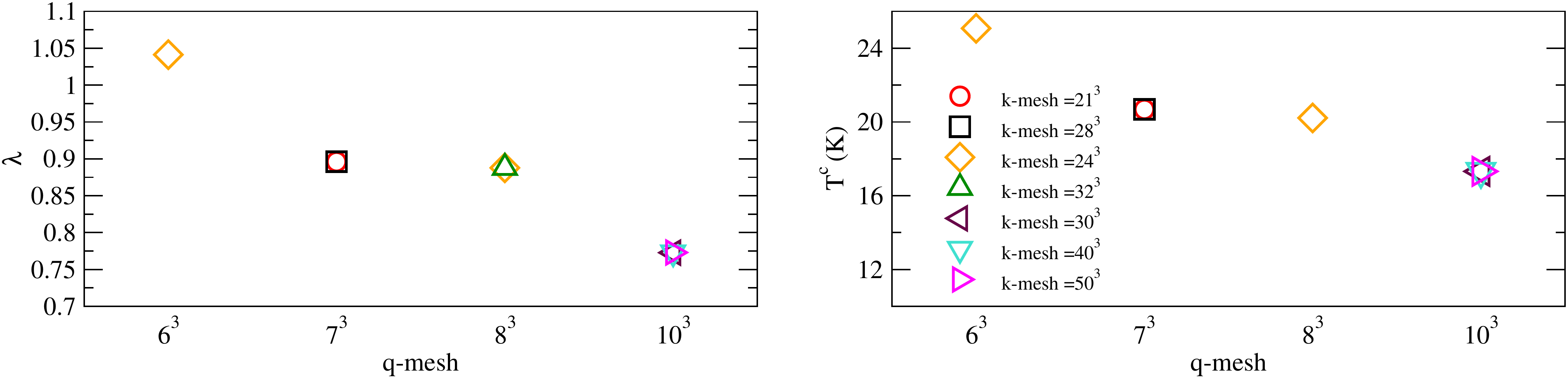}
\caption{Convergence tests for electron-phonon coupling constant, $\lambda$ and $T_c$ with fixed Gaussian broadening 0.005 (Ry) for different $\mathbf{k}$-mesh and $\mathbf{q}$-mesh. }
\label{fig:epi_convergence}
\end{figure*}

\begin{figure*}[!b]
\includegraphics[width=0.94\textwidth]{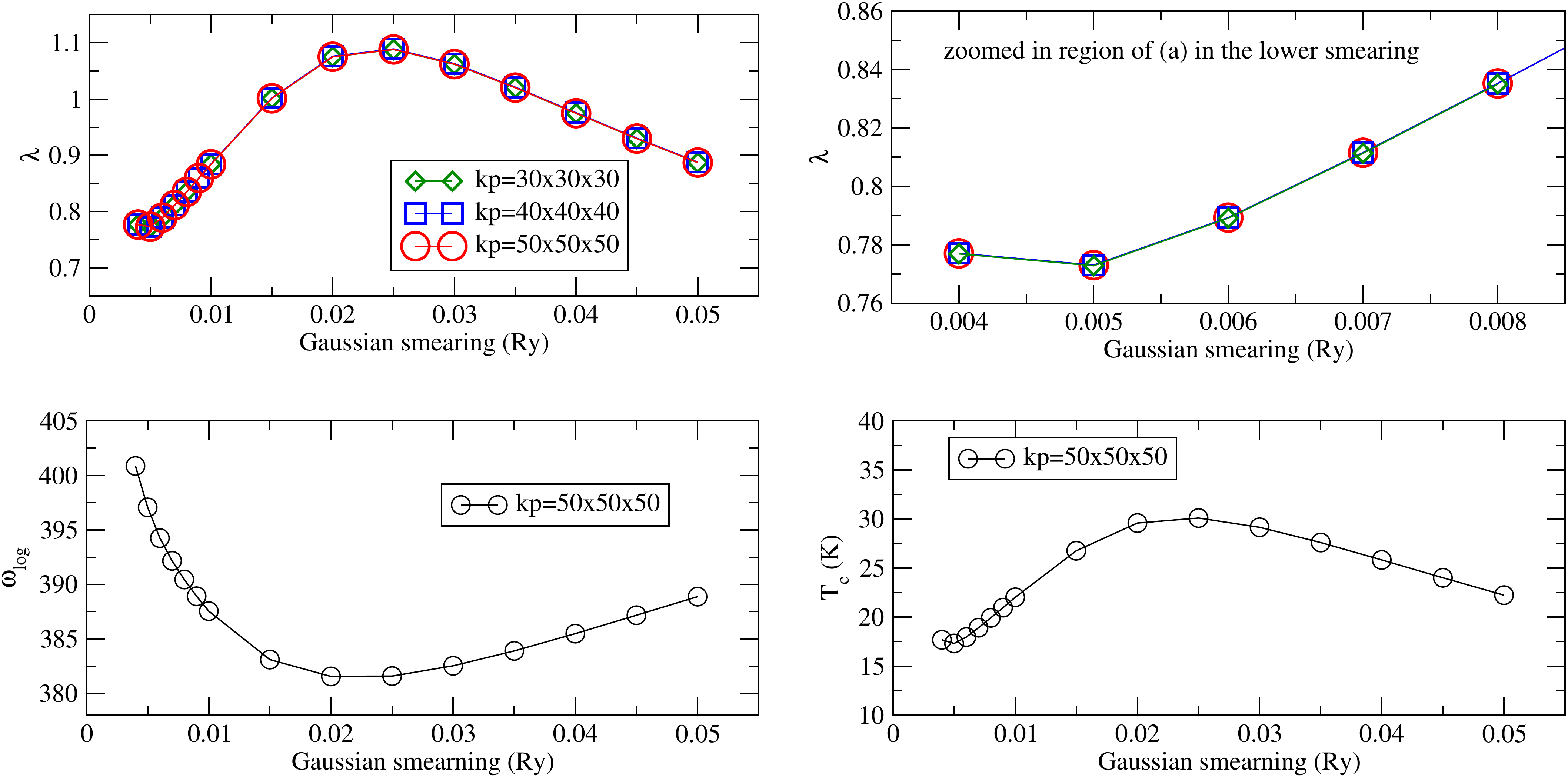}
\caption{Convergence tests for electron-phonon coupling constant, $\lambda$, $\omega_{log}$ and $T_c$ with Gaussian smearing value and $\mathbf{k}$-mesh, for fixed $\mathbf{q}$-mesh=$10\times 10\times10$.  }
\label{fig:epi_q10}
\end{figure*}

As mentioned in the main text and shown in Fig.~\ref{fig:ph_disp_pressure}, the phonon dispersions show imaginary frequencies over the entire pressure range (only two pressure values are taken for demonstration) when we keep the structures at the experimental lattice constants~\cite{pei2021pressureinduced}.  As the calculation of electron-phonon coupling generally requires a very good convergence check of $\mathbf{k}$-mesh and $\mathbf{q}$-meshes, we had performed the convergence tests for the relaxed (calculated) structure at 100 GPa as shown in Figs~\ref{fig:epi_convergence} and ~\ref{fig:epi_q10}. We take the converged values of $\mathbf{k}$-mesh $\mathbf{q}$-mesh to be $30 \times 30 \times 30$  and $10 \times 10 \times 10$, respectively and gaussian smearing value is $0.004$ Ry.

We have, however, estimated the values of EPC constants, $\lambda$, and critical temperature, $T_c$, in  experimental structures at 60.77 and 100 GPa
 by removing the imaginary frequencies which give the lower bound of $\lambda$. We found  $T_c$ to be lowered by $~$1 K from 100 GPa to 60.77 GPa, giving rise to a slope of $0.018$ in the pressure-temperature plot. This is motivated us to draw the `EPC-only' contribution (black line) as linear in Fig. 1 of the main text. Note that in our case, we refrain from reporting the actual calculated values of $T_c$ (only diffrences are considered), which could be obtained by McMillan-Allen-Dynes formula~\cite{McMillan, Allen-Dynes},  due to the following reason: As our estimate of $\lambda$ of experimental structures were done with the removal of one acoustic phonon, it gives larger logarithmic average of phonon frequencies, $\omega_{log}$,  and hence larger $T_c$ values. For the correct estimate, however, one should perhaps consider the phonon frequencies of the supercells (as described in section IV), suggested by the phonon instabilities. As our purpose here is to obtain a qualitative understanding, we avoided performing such calculations.

\section{Charge density wave (CDW) states }

Phonon dispersions of the experimental structures~\cite{pei2021pressureinduced} show dynamic instabilities through the appearance of an imaginary longitudinal acoustic mode along the $c$-direction which is plotted as negative frequencies in Fig.~\ref{fig:ph_disp_pressure}. Note that these instabilities appear at all pressures for the experimental structures. Below we show that \crb\ under pressure develops a charge density wave (CDW) state.

We have performed a supercell calculation at 100 GPa following the phonons instability perpendicular to the hexagonal boron plane. Relaxation of the supercell ($\sqrt{3}a \times 3a \times 2c$) gives rise to a body-centered orthorhombic structure, where one third of the Cr atom is puckered (by 0.085 \AA) along the $c$-direction due to application of pressure, in addition to the appearance of  nonuniform Cr bonds (which differ by maximum amount of 0.024 \AA). The conventional cell is 4.5 meV/\crb\ higher in energy than the supercell, suggesting a CDW state with  $q={1/3,1/2,1/2}$. However, zero-point vibrational fluctuations may stabilize the conventional structure.

\section{Two band toy model}

To illustrate the effect of pair breaking from spin-fluctuation pairing processes in combination with electron-phonon pairing, we consider a simple two band toy model and cast the linearized gap equation, Eq.~(\ref{eqn:gapeqn}) into a $2\times 2$ matrix equation
\begin{equation}
 -\underline{V}\; \underline{\rho} \;g_\pm =\lambda_\pm g_\pm
\end{equation}
where $\underline{V}=\underline{V^{\text{SF}}}+\underline{V^{\text{el-ph}}}$ is the (averaged over the Fermi surface) pairing interaction from spin-
fluctuations (SF) and from the electron-phonon (el-ph) mechanism and $\underline{\rho}$ is a (diagonal) matrix with the partial densities of states of the two bands. $\lambda_\pm$ are the two eigenvalues to the two eigenvectors $g_\pm$. For simplicity, we assume two bands with identical density of states,  $\underline{\rho}=\rho_0 \delta_{ij}$ and parametrize the pairing interactions (including the density of states prefactor as follows). The electron-phonon pairing in CrB$_2$ is dominated by small momentum transfer $\bf q$, thus we assume it to be intra-band only,
\begin{equation}
 \underline{V^{\text{el-ph}}}=\left(\begin{array}{cc}
                                     \alpha&\epsilon\\
                                     \epsilon& \alpha\\
                                    \end{array}
\right)\; ,\label{eq_Velph}
\end{equation}
i.e. $|\alpha|\gg |\epsilon|$.
The spin-fluctuation part has intra-band and inter-band contributions,
\begin{equation}
 \underline{V^{\text{SF}}}=\left(\begin{array}{cc}
                                     \beta&\delta\\
                                     \delta& \beta\\
                                    \end{array}
\right)\; .\label{eq_Vsf}
\end{equation}
The spin-fluctuation pairing has the following dependence on the susceptibility $\Gamma\sim U + U^ 2\chi({\bf q})^{\text{RPA}}=U + U^ 2\chi({\bf q})^{0}/(1-U\chi({\bf q})^{0})$ where $\bf q$ is either the sum or difference of the scattering vectors connecting the two Fermi surfaces (when projecting the pairing interaction into the singlet channel).
With the expected behavior of the susceptibility close to a nesting condition of $\chi^0({\bf q})=\gamma/(1+({\bf q}-{\bf q}_c)^2\xi^2)$ where ${\bf q}_c$ is the nesting vector, $\xi$ the correlation length of the spin-fluctuations and $\gamma$ proportional to the (inverse) bandwidth $W$ of the electronic structure.
We are assuming further the following linear dependence of the bandwidth $W(p)=W_{\text{AFM}} (1+x(p-p_c))$, where $W_{\text{AFM}}$ is the bandwidth at the antiferromagnetic instability, $p_c$ is the corresponding critical pressure and $x=...$ the coefficient from the Taylor expansion.
We can then rewrite the spin fluctuation pairing coefficients as $\beta=\beta_0/(1-\alpha_1 + x(p-p_c))$ and $\delta=\beta_0/(1-\alpha_2 + x(p-p_c))$ where the parameter $\alpha_i$ controls the closeness to the nesting condition, i.e. $\alpha_i=1$ is perfect nesting of all $k$-points on the Fermi surfaces and $\alpha_i=0$ no nesting such that the susceptibility is just given by the background from the density of states at the Fermi level.
\begin{figure}[tb]
\includegraphics[width=0.5\linewidth]{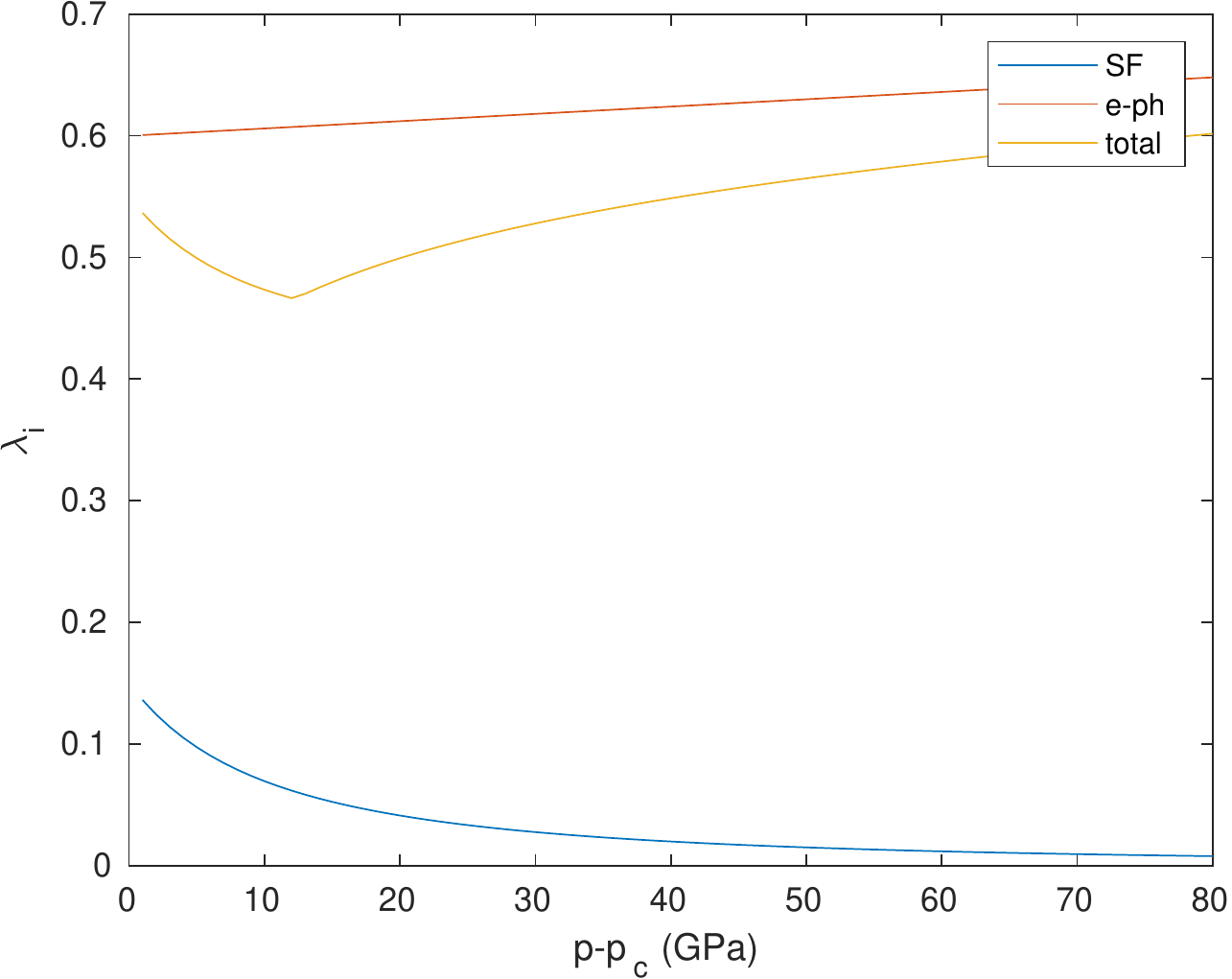}
\caption{Cooperation and competition of spin-fluctuation pairing and electron-phonon pairing. Plot of the largest eigenvalues for the two band model showing the transition from an $s_{\pm}$ state with sizable spin-fluctuation pairing close to the antiferromagnetic instability at $p_c$ to an ordinary $s_{++}$ state where pair-breaking effects from spin-fluctuations are still sizeable.}
\label{fig_lambda_phen}
\end{figure}

Looking at the two parts of the pairing interaction separately, it is evident that the matrix in Eq.~(\ref{eq_Velph}) yields two positive eigenvalues with eigenvectors $g_+=(1,1)/\sqrt{2}$ and $g_-=(1,-1)/\sqrt{2}$ where the first has larger eigenvalue and corresponds to the (usual) $s_{++}$ pairing state. The matrix in Eq.~(\ref{eq_Vsf}) has the same eigenvectors, where only $g_-$ yields a positive eigenvalue. In Fig. \ref{fig_lambda_phen} we show the overall behavior of the individual eigenvalues and the combined pairing interaction that leads to a significantly reduced eigenvalue and thus $T_\mathrm{c}$ close to the antiferromagnetic instability and a change of the pairing state from $s_{\pm}$ to $s_{++}$ as pressure increases.

\section{Functional Renormalization Group (fRG) calculations }
The inherent three dimensionality and the importance of multi-orbital features in CrB$_2$ present significant challenges for any many body calculation in this system. While these conditions pose a significant challenge for our RPA analysis, they become insurmountable for functional renormalization group (fRG) calculations. Since the fRG extends the analysis of the effective interaction beyond the Cooper channel, an at least approximate treatment of the full momentum dependence for the two-particle vertex function is a prerequisite. In the most modern formulation of the fRG, the truncated unity (TU) approximation, an RPA result guided resolution would still necessitate an effective vertex description with over 20 billion parameters. Within the fRG this equates to the solution of a coupled system of 20 billion integro-differential equations which is far beyond the current cutting edge of possibilities. The problem is further complicated by the TU-approximations incompatibility with the use of a natural basis for an efficient implementation of symmetries~\cite{Beyer2022}.

\end{document}